\PassOptionsToPackage{unicode}{hyperref}
\PassOptionsToPackage{hyphens}{url}
\PassOptionsToPackage{dvipsnames,svgnames*,x11names*}{xcolor}
\documentclass[
  british,
]{article}
\usepackage{lmodern}
\usepackage{setspace}
\usepackage{amssymb,amsmath}
\usepackage{ifxetex,ifluatex}
\ifnum 0\ifxetex 1\fi\ifluatex 1\fi=0 
  \usepackage[T1]{fontenc}
  \usepackage[utf8]{inputenc}
  \usepackage{textcomp} 
\else 
  \usepackage{unicode-math}
  \defaultfontfeatures{Scale=MatchLowercase}
  \defaultfontfeatures[\rmfamily]{Ligatures=TeX,Scale=1}
\fi
\IfFileExists{upquote.sty}{\usepackage{upquote}}{}
\IfFileExists{microtype.sty}{
  \usepackage[]{microtype}
  \UseMicrotypeSet[protrusion]{basicmath} 
}{}
\makeatletter
\@ifundefined{KOMAClassName}{
  \IfFileExists{parskip.sty}{%
    \usepackage{parskip}
  }{
    \setlength{\parindent}{0pt}
    \setlength{\parskip}{6pt plus 2pt minus 1pt}}
}{
  \KOMAoptions{parskip=half}}
\makeatother
\usepackage{xcolor}
\IfFileExists{xurl.sty}{\usepackage{xurl}}{} 
\IfFileExists{bookmark.sty}{\usepackage{bookmark}}{\usepackage{hyperref}}
\hypersetup{
  pdftitle={No Deal: Investigating the Influence of Restricted Access to Elsevier Journals on German Researchers' Publishing and Citing Behaviours},
  pdflang={en-GB},
  colorlinks=true,
  linkcolor=black,
  filecolor=Maroon,
  citecolor=Blue,
  urlcolor=blue,
  pdfcreator={LaTeX via pandoc}}
\usepackage[margin=1in]{geometry}
\usepackage{longtable,booktabs}
\usepackage{etoolbox}
\makeatletter
\patchcmd\longtable{\par}{\if@noskipsec\mbox{}\fi\par}{}{}
\makeatother
\IfFileExists{footnotehyper.sty}{\usepackage{footnotehyper}}{\usepackage{footnote}}
\makesavenoteenv{longtable}
\usepackage{graphicx}
\makeatletter
\def\maxwidth{\ifdim\Gin@nat@width>\linewidth\linewidth\else\Gin@nat@width\fi}
\def\maxheight{\ifdim\Gin@nat@height>\textheight\textheight\else\Gin@nat@height\fi}
\makeatother
\setkeys{Gin}{width=\maxwidth,height=\maxheight,keepaspectratio}
\makeatletter
\def\fps@figure{htbp}
\makeatother
\providecommand{\tightlist}{%
  \setlength{\itemsep}{0pt}\setlength{\parskip}{0pt}}
\ifxetex
  \usepackage{polyglossia}
  \setmainlanguage[variant=british]{english}
\else
  \usepackage[shorthands=off,main=british]{babel}
\fi
\ifluatex
  \usepackage{selnolig}  
\fi
\newlength{\cslhangindent}
\newlength{\csllabelwidth}
\newenvironment{CSLReferences}[3] 
 {
  \setlength{\parindent}{0pt}
  \ifodd #1 \everypar{\setlength{\hangindent}{\cslhangindent}}\ignorespaces\fi
  \ifnum #2 > 0
  \setlength{\parskip}{#3\baselineskip}
  \fi
 }%
 {}
\usepackage{calc} 

\title{\vspace{-2em}No Deal: Investigating the Influence of Restricted Access to Elsevier Journals on German Researchers' Publishing and Citing Behaviours\vspace{-3em}}
\date{}

\begin{document}
\maketitle

\newcommand{\orcid}{%
  \begingroup\normalfont
  \includegraphics[height=6px]{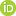}%
  \endgroup
}
Nicholas Fraser\textsuperscript{1,*} (\orcid{} \href{https://orcid.org/0000-0002-7582-6339}{\color{black}{0000-0002-7582-6339}}), Anne Hobert\textsuperscript{2} (\orcid{} \href{https://orcid.org/0000-0003-2429-2995}{\color{black}{0000-0003-2429-2995}}), Najko Jahn\textsuperscript{2} (\orcid{} \href{https://orcid.org/0000-0001-5105-1463}{\color{black}{0000-0001-5105-1463}}), Philipp Mayr\textsuperscript{3} (\orcid{} \href{https://orcid.org/0000-0002-6656-1658}{\color{black}{0000-0002-6656-1658}}), Isabella Peters\textsuperscript{1,4} (\orcid{} \href{https://orcid.org/0000-0001-5840-0806}{\color{black}{0000-0001-5840-0806}}). \\

\textsuperscript{1} ZBW -- Leibniz Information Centre for Economics, Kiel, Germany \\
\textsuperscript{2} Göttingen State and University Library, University of Göttingen, Göttingen, Germany \\
\textsuperscript{3} GESIS -- Leibniz Institute for the Social Sciences, Cologne, Germany \\
\textsuperscript{4} Kiel University, Kiel, Germany \\

\textsuperscript{*} Correspondence: \href{mailto:n.fraser@zbw.eu}{\color{black}{n.fraser@zbw.eu}} 

\setstretch{1.5}
\hypertarget{abstract}{%
\subsection{Abstract}\label{abstract}}

In 2014, a union of German research organisations established Projekt DEAL, a national-level project to negotiate licensing agreements with large scientific publishers. Negotiations between DEAL and Elsevier began in 2016, and broke down without a successful agreement in 2018; in this time, around 200 German research institutions cancelled their license agreements with Elsevier, leading Elsevier to restrict journal access at those institutions from July 2018 onwards. We investigated the effect of these access restrictions on researchers' publishing and citing behaviours from a bibliometric perspective, using a dataset of \textasciitilde410,000 articles published by researchers at the affected DEAL institutions between 2012-2020. We further investigated these effects with respect to the timing of contract cancellations with Elsevier, research disciplines, collaboration patterns, and article open-access status. We find evidence for a decrease in Elsevier's market share of articles from DEAL institutions, from a peak of 25.3\% in 2015 to 20.6\% in 2020, with the largest year-on-year market share decreases occurring in 2019 (-1.1\%) and 2020 (-1.6\%) following the implementation of access restrictions. We also observe year-on-year decreases in the proportion of citations made from articles published by authors at DEAL institutions to articles in Elsevier journals post-2018, although the decrease is smaller (-0.4\% in 2019 and -0.6\% in 2020) than changes in publishing volume. We conclude that Elsevier access restrictions have led to some reduced willingness of researchers at DEAL institutions to publish their research in Elsevier journals, but that researchers are not strongly affected in their ability to cite Elsevier articles, with the implication that researchers use a variety of other methods (e.g.~interlibrary loans, sharing between colleagues, or ``shadow libraries'') to access scientific literature.

\hypertarget{declarations}{%
\subsection{Declarations}\label{declarations}}

\hypertarget{author-contributions}{%
\paragraph{Author Contributions}\label{author-contributions}}

Conceptualisation: NF; Data curation: NF, AH; Formal analysis: NF; Funding acquisition: PM, IP; Investigation: NF; Methodology: NF, AH; Project administration: PM, IP; Software: NF; Supervision: PM, IP; Validation: NF; Visualisation: NF; Writing - original draft: NF; Writing - review and editing: NF, AH, NJ, PM, IP.

\hypertarget{competing-interests}{%
\paragraph{Competing interests}\label{competing-interests}}

The authors declare no competing interests.

\hypertarget{acknowledgments}{%
\paragraph{Acknowledgments}\label{acknowledgments}}

This work was supported by the German Federal Ministry of Education and Research within the
funding stream ``Quantitative research on the science sector'', projects OASE (grant numbers 01PU17005A and 01PU17005B) and OAUNI (grant numbers 01PU17023A and 01PU17023B). We are grateful to Dimensions (\url{https://www.dimensions.ai/}) for providing API access through their scientometric research access program.

\hypertarget{data-and-code-availability}{%
\paragraph{Data and code availability}\label{data-and-code-availability}}

Aggregated datasets presented in this manuscript, as well as all code used for the data extraction, analysis, and manuscript preparation are available on GitHub (\url{https://github.com/nicholasmfraser/Projekt_DEAL}) and archived on Zenodo (\url{https://doi.org/10.5281/zenodo.4771576}).

\pagebreak

\hypertarget{introduction}{%
\subsection{Introduction}\label{introduction}}

In 2014 the \href{https://wissenschaftsfreiheit.de/}{Alliance of Science Organisations} (\emph{Allianz der Wissenschaftsorganisationen}; AWO), a union of the majority of German research organisations, established a national-level project named \href{https://www.projekt-deal.de}{``Projekt DEAL''} (herein referred to as ``DEAL''), to negotiate licensing agreements for access to electronic journals of large scientific publishers. The key objectives of DEAL are:

\begin{enumerate}
\def\labelenumi{\arabic{enumi}.}
\tightlist
\item
  To receive permanent, full-text access to the entire journal portfolio of the selected publishers.
\item
  To make all articles published by authors at German institutions automatically Open Access (OA).
\item
  To secure reasonable pricing according to a simple, future-oriented model based on publication volumes.
\end{enumerate}

To date, DEAL negotiations have centred on three major publishers: Elsevier, Springer Nature and Wiley. Between 2012 and 2020, these three publishers were collectively responsible for publishing \textasciitilde53\% of scientific articles with at least a single author from a German research institution (Elsevier \textasciitilde21\%, Springer Nature \textasciitilde21\%, Wiley \textasciitilde12\%; data according to \href{https://www.dimensions.ai/}{Dimensions}). Negotiations between DEAL and Elsevier officially began in 2016, with Springer Nature and Wiley negotiations beginning a year later in 2017. In January 2019, DEAL announced the signing of an agreement with Wiley, fulfilling the defined negotiating objectives by allowing full access to Wiley's portfolio of journals for institutions represented by DEAL (herein referred to as ``DEAL institutions''), and automatic OA publishing of articles from corresponding authors at DEAL institutions under Creative Commons (CC) licenses, for an annual fee equal to €2,750 per published article (\href{https://doi.org/10.17617/2.3027595}{Sander et al., 2019}). In January 2020 a similar agreement between DEAL and Springer Nature was signed, including the same per-article fee equal to €2,750 (\href{https://doi.org/10.17617/2.3174351}{Kieselbach, 2020}).

Whilst negotiations with Wiley and Springer Nature have now successfully concluded in agreements, negotiations with Elsevier remain unresolved. At the end of 2016, \textasciitilde70 German institutions\footnote{Archived webpage: \url{https://web.archive.org/web/20191212094238/https://www.projekt-deal.de/vertragskuendigungen_elsevier/}} chose not to renew their contracts with Elsevier, leading to Elsevier restricting access to new journal issues at those institutions (and also restricting access to back-catalogues at some institutions) from the beginning of 2017 (\href{https://doi.org/10.1126/science.355.6320.17}{Vogel, 2017a}), although access was restored six weeks later (\href{https://doi.org/10.1126/science.aal0753}{Vogel, 2017b}). At the end of 2017, a further \textasciitilde110 German institutions\footnote{Archived webpage: \url{https://web.archive.org/web/20191212094842/https://www.projekt-deal.de/vertragskundigungen-elsevier-2017/}} decided not to renew their contracts with Elsevier, and at the beginning of July 2018, the \href{https://www.hrk.de/}{German Rectors' Conference} (\emph{Hochschulrektorenkonferenz}; HRK), who are leading negotiations on behalf of AWO, \href{https://www.hrk.de/presse/pressemitteilungen/pressemitteilung/meldung/verhandlungen-von-deal-und-elsevier-elsevier-forderungen-sind-fuer-die-wissenschaft-inakzeptabel-440/}{announced} the breakdown and cancellation of all ongoing negotiations with Elsevier. In July 2018, authors at institutions which had cancelled their contracts with Elsevier had their access to new journal issues completely cut-off (\href{https://doi.org/10.1038/d41586-018-05754-1}{Else, 2018}). A further \textasciitilde25 institutions, including the Max Plank Society and Fraunhofer Society\footnote{Archived webpage: \url{https://web.archive.org/web/20191114074949/https://www.projekt-deal.de/vertragskundigungen-elsevier-2018/}} did not renew their contracts with Elsevier at the end of 2018, bringing the total number of institutions without a contract with Elsevier to more than 200.

As a provider of a large proportion of research published and cited by researchers at German institutions, restricted access to Elsevier's article collections may have measurable effects on their publication and citation behaviour. Attempts to quantify such effects have already been made through recent survey approaches: A survey commissioned by Elsevier\footnote{Archived webpage: \url{https://web.archive.org/web/20210427160017/https://resources.mynewsdesk.com/image/upload/fl_attachment/sgmm6zhwdmam4ys7evvf}} found that 61\% of researchers at German institutions agreed or strongly agreed that losing access made their research activities less efficient, whilst 54\% agreed or strongly agreed that losing access delayed the speed that they produce their research output. A separate survey of 384 researchers at the Faculty of Medicine of the University of Münster showed an overall similar sentiment, with 66\% of researchers reporting that they now require more time to acquire literature and 46\% of researchers reporting that losing access was a competitive disadvantage, yet only 29\% of researchers reported that they would no longer write or review articles for Elsevier journals\footnote{Archived webpage: \url{https://web.archive.org/web/20210429122105/https://www.uni-muenster.de/ZBMed/aktuelles/27850}}. Following the implementation of access restrictions, a number of researchers also resigned from their positions on editorial boards of Elsevier journals\footnote{Archived webpage: \url{https://web.archive.org/web/20210427081142/https://www.projekt-deal.de/aktuelles-zu-elsevier/}} (\href{https://doi.org/10.1126/science.aar2142}{Vogel, 2017c}).

The situation in Germany is not unique and breakdowns in negotiations between library consortia and Elsevier have been reported elsewhere. In Sweden, a number of universities, research institutes and government agencies were cut off from Elsevier journals between mid-2018 and the end of 2019 due to a breakdown in negotiations between Elsevier and the \href{https://www.kb.se/samverkan-och-utveckling/oppen-tillgang-och-bibsamkonsortiet.html}{Bibsam Consortium} (the national-level license negotiating body for Sweden). An \href{https://www.kb.se/download/18.a9bd5bf1707b0801cd15e/1582893792629/Bibsam-Elsevier-2020-2022-tobepublished-titlelistexcluded.pdf}{agreement} was eventually signed between Bibsam and Elsevier at the end of 2019, to take effect from 1st January 2020. A survey of 4,221 Swedish researchers carried out by Bibsam during the time period that Elsevier journals were inaccessible found that 51\% of respondents were negatively affected in their desire to publish with Elsevier, and 54\% had their work negatively impacted (\href{http://doi.org/10.1629/uksg.507}{Olson et al., 2020}). The University of California (UC) system also had access to Elsevier journals restricted following the suspension of negotiations in 2019. A poll of \textasciitilde7,300 UC affiliates (including faculty, graduate students, undergraduates, postdocs and other staff) during this time period found that 33\% of respondents reported a significant impact of the loss of access, most greatly from the health sciences (52\% reported a significant impact), but only 15\% reported that it would affect their decision to publish in Elsevier journals\footnote{Archived webpage: \url{https://web.archive.org/web/20210518094220/https://www.library.ucsb.edu/uc-and-elsevier}}. In March 2021, the University of California \href{https://osc.universityofcalifornia.edu/2021/03/uc-secures-landmark-oa-deal-with-worlds-largest-scientific-publisher/}{announced} a 4-year agreement with Elsevier, which, for the first time, also covers OA publication in the Cell Press and Lancet family of journals.

In parallel to these library-led negotiations, other researcher-led protests against the business practices of Elsevier have occurred in recent years. \href{http://thecostofknowledge.com/}{The Cost of Knowledge} is a campaign launched by mathematician Timothy Gowers in 2012, asking researchers to sign a statement that they would refrain from publishing, refereeing or doing editorial work for Elsevier. To date, more than 18,000 researchers have publicly signed the statement. However, an analysis of the signatories in 2016 found that 38\% of those who committed to not publish in Elsevier journals actually published papers in Elsevier journals following their commitment (\href{https://doi.org/10.3389/frma.2016.00007}{Heyman et al., 2016}). Other protests have been made in the form of editorial board resignations; in 2015 the entire editorial board of \href{https://www.journals.elsevier.com/lingua}{Lingua}, an Elsevier journal, resigned and formed a new journal named \href{https://www.glossa-journal.org/}{Glossa} with a different publisher; a similar situation unfolded in 2019 when the editorial board of the \href{https://www.journals.elsevier.com/journal-of-informetrics}{Journal of Informetrics} resigned and formed a new journal named \href{https://direct.mit.edu/qss}{Quantitative Science Studies}. However, in both cases, the existing Elsevier journals continued to operate and publish new journal issues with newly-established editorial boards.

In this study, we aim to investigate how restricted access to Elsevier journals at DEAL institutions in Germany had a direct effect on researchers' publishing and citing behaviours. Despite suggestions from surveys that researchers were negatively affected in their desire to publish in Elsevier journals and negatively impacted in their daily research activities, to our knowledge no empirical evidence suggesting that this has actually occurred has been established. Specifically, we aim to answer the following research questions:

\begin{enumerate}
\def\labelenumi{\arabic{enumi}.}
\tightlist
\item
  Did restricted access to Elsevier journals at DEAL institutions result in a change in researchers' publishing behaviour?
\item
  Did restricted access to Elsevier journals at DEAL institutions result in a change in researchers' citing behaviour?
\end{enumerate}

For both of these research questions, we also consider variability with respect to the timing of contract cancellations of individual institutions, research disciplines, collaboration patterns, and article OA status.

\hypertarget{methods}{%
\subsection{Methods}\label{methods}}

\hypertarget{data-sources}{%
\subsubsection{Data Sources}\label{data-sources}}

\hypertarget{deal-institutions}{%
\paragraph{DEAL institutions}\label{deal-institutions}}

We collected names and contract expiration dates of 210 German universities, research institutions, higher education institutions and regional libraries that had their access to Elsevier articles restricted as part of the DEAL negotiations, using publicly available lists of institutions that cancelled their contracts with Elsevier in 2016\footnote{Archived webpage: \url{https://web.archive.org/web/20191212094238/https://www.projekt-deal.de/vertragskuendigungen_elsevier/}}, 2017\footnote{Archived webpage: \url{https://web.archive.org/web/20191212094842/https://www.projekt-deal.de/vertragskundigungen-elsevier-2017/}} and 2018\footnote{Archived webpage: \url{https://web.archive.org/web/20191114074949/https://www.projekt-deal.de/vertragskundigungen-elsevier-2018/}} available on the \href{https://www.projekt-deal.de/}{DEAL website}. We manually mapped institutions to identifiers in the \href{https://www.grid.ac/}{Global Research Identifier Database} (GRID), using the available search interface on the GRID website. Of the 210 institution names provided, 209 were matched to a GRID identifier, with the exception of ``HS Villingen-Schwenningen'', for which we were unable to unambiguously determine the correct GRID identifier. The list extracted from the Project DEAL website contained 10 individual records for each campus associated with the Baden-Wuerttemberg Cooperative State University (``Duale Hochschule Baden-Württemberg / DHBW''), but all campuses are collectively associated with a single identifier in the GRID database (``grid.449295.7''). The Max Planck Society and Fraunhofer Society were listed individually on the DEAL website, but both are umbrella associations that consist of a number of individual research institutions. Thus, we also extracted the GRID information for all their constituents, according to ``parent-child'' relationship information stored in GRID. The Helmholtz Association and Leibniz Association are similar umbrella associations, but the lists from the DEAL website contained the names of individual Helmholtz Association and Leibniz Association institutions, thus we limited the dataset to those contained directly in the list and did not extract information for all other constituent institutions. A limitation of this approach is that we rely solely on publicly available information of contract negotiations and cancellations; nonetheless, we attempted to verify the information contained on the DEAL website by manually searching for press releases or other informational web pages issued or maintained by the individual institutions that referred to restricted access to Elsevier articles. Of the original 210 institutions on the list, we found relevant information for 121 institutions; the information we discovered showed strong concordance with the information contained on the DEAL website (e.g.~in terms of contract status and timing of restricted access) and thus we are confident in the quality of the publicly-available information.

\hypertarget{article-and-author-metadata}{%
\paragraph{Article and author metadata}\label{article-and-author-metadata}}

\begin{figure}

{\centering \includegraphics{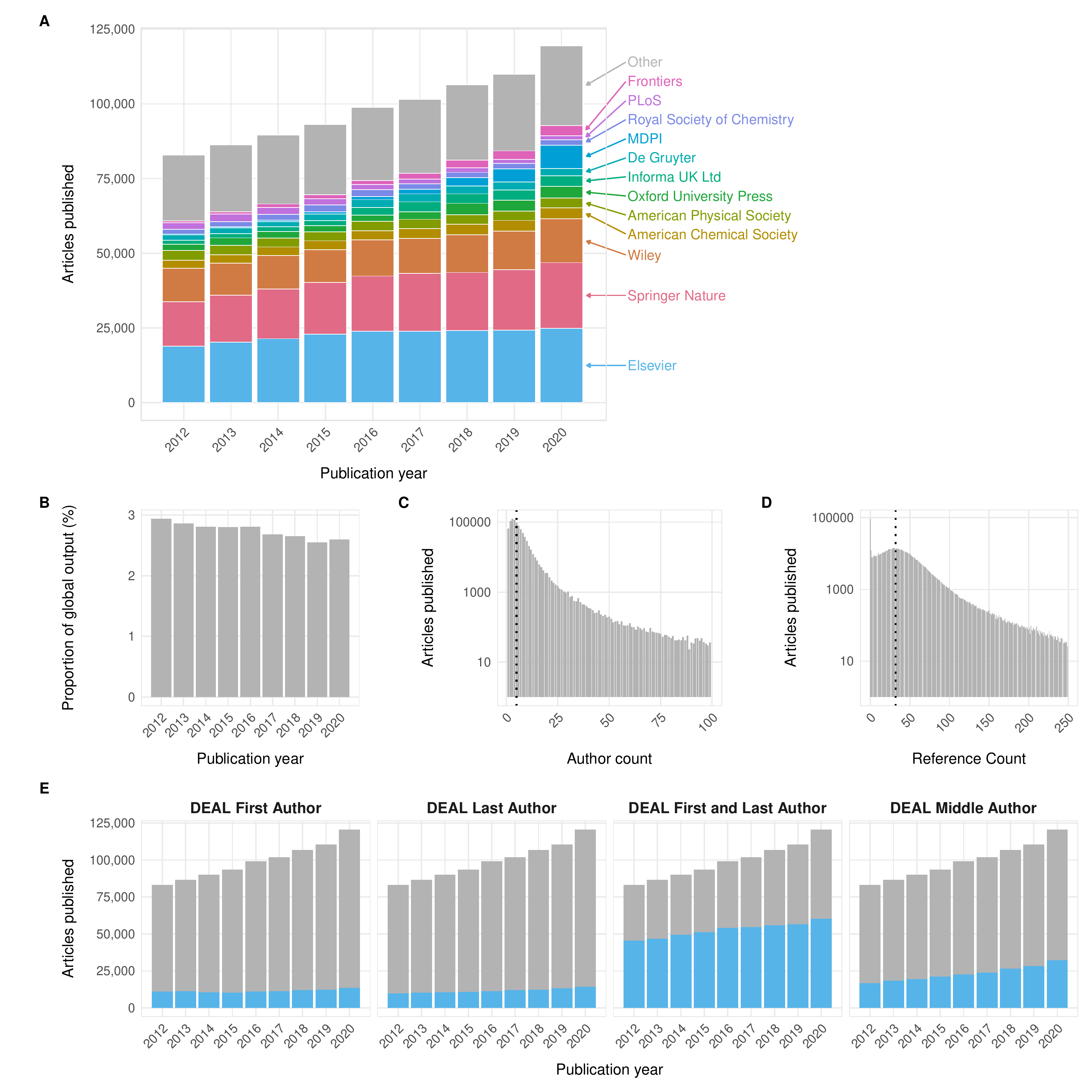} 

}

\caption{Properties of our initial dataset of articles published by DEAL researchers extracted from Dimensions. (A) Number of articles published per year and by publisher (top-12 publishers by total publication volume shown). (B) Proportion of articles in our dataset as a proportion of total global output per year (according to total number of articles indexed in Dimensions). (C) Distribution of number of authors per article (only articles with \textless100 authors are shown). Note y-axis is on a log-scale. The dotted line refers to median number of authors per article (5). (D) Distribution of number of references per article (only articles with \textless250 references are shown). Note y-axis is on a log-scale. The dotted line refers to median number of references per article (32). (E) Proportion of articles by authorship type: ``DEAL First Author'' refers to articles where the first author is from a DEAL institution but the last author is not, ``DEAL Last Author'' where the last author is from a DEAL institution but the first author is not, ``DEAL First and Last Author'' where both the first and last authors are from DEAL institutions, and ``DEAL Middle Author'' where neither the first nor last authors are from DEAL institutions.}\label{fig:items-overview}
\end{figure}

Article and author metadata used in this study were derived from three main bibliometric data sources: \href{https://www.dimensions.ai/}{Dimensions}, \href{http://crossref.org/}{Crossref} and \href{https://unpaywall.org/}{Unpaywall}. Initially, we retrieved article DOIs, complete author and affiliation details, fields of research, and reference lists (DOI-DOI links) for all articles with at least a single author from a DEAL institution, via the Dimensions Analytics API. These data were retrieved in the first two weeks of April 2021. Articles were limited to those with a publication date in years 2012 to 2020, and to ``article'' publication types. As the Dimensions Analytics API only allows a maximum of 50,000 records to be returned in a single query, we queried iteratively through each year and individual DEAL institution, using the associated GRID identifier, and extracted details of all articles that included an author at the respective institution. In a final step we combined all article records together and removed duplicates (e.g.~where an article had authors from multiple DEAL institutions). Following these steps we created a set of 892,169 unique articles (overview of distribution over time and by publisher in Figure \ref{fig:items-overview}A), representing \textasciitilde2.5-3\% of total global output over the same time period (Figure \ref{fig:items-overview}B).

Figure \ref{fig:items-overview}C shows the distribution of the number of authors per article in our original dataset. The figure shows that a high proportion of articles contain a large number of authors. In cases of articles written by large teams or consortia, the contribution of DEAL authors to the writing of the article or subsequent publication strategy may be small. Figure \ref{fig:items-overview}E shows the proportion of articles in our original dataset in each year further subdivided by authorship types: ``DEAL First Author'' refers to articles where the first author is from a DEAL institution but the last author is not, ``DEAL Last Author'' where the last author is from a DEAL institution but the first author is not, ``DEAL First and Last Author'' where both the first and last authors are from DEAL institutions, and ``DEAL Middle Author'' where neither the first nor last authors are from DEAL institutions. These results show that the majority of articles in our dataset have a first or last author (or both) from DEAL institutions, yet there exist a number of articles where DEAL authors are only included as middle authors. Although practices for the assignment of author order are neither clear nor consistent across disciplines (\href{https://doi.org/10.1087/20150211}{Brand et al., 2015}), for the purposes of this study we make the assumption that the publication strategy for the article is primarily determined by either the first or last author of an article. Thus, as our study aims to focus on the direct behaviour of researchers at DEAL institutions, we subsequently limited our dataset to articles with a first AND last author from a DEAL institution (i.e.~the group ``DEAL First and Last Author'' in Figure \ref{fig:items-overview}E)\footnote{An alternative approach here would have been to use ``corresponding author'' information rather than author positions; however, an analysis of a subset of articles from 10 randomly selected institutions (N articles = 6,513) revealed that only 57\% of articles contained any corresponding author information, and coverage may be biased towards certain publishers. Of articles in this subset, we found that in 99\% of cases where the first and last author were from a DEAL institution, that the corresponding author was also from a DEAL institution; although we may remove more articles from the analysis in our approach, we do not introduce any potential biases that may result from using incomplete corresponding author information.}.

For each article retrieved from Dimensions, we also retrieved and parsed a complete list of references (total number of DOI-DOI reference links: 33,652,274). An overview of the distribution of references per article for our original dataset is shown in Figure \ref{fig:items-overview}D. We observed that an anomalously high number of articles in this distribution contained either zero references or a single reference. A manual investigation on a small random sample of these articles revealed that articles with zero references often represent diverse types of editorial content (e.g.~corrections, errata, tables of content) which is indexed in Dimensions as ``article'' types. Articles with a single reference often represent abstracts which are linked to a single journal article (see, for an example, abstracts published by the journal \href{https://onlinelibrary.wiley.com/journal/15222667}{ChemInform}). In a small number of cases, articles that did not contain references in our dataset did in fact contain a reference list on the journal page - highlighting a potential weakness in Dimensions as a data source. However, as a broad generalisation, we conclude that articles containing zero references or a single reference do not represent ``true'' research articles, which is the focus of this study\footnote{Unpaywall also provide an indicator of potential editorial content (defined in their data schema as ``is\_paratext'') which \href{https://support.unpaywall.org/support/solutions/articles/44001894783-what-does-is-paratext-mean-in-the-api-}{is based on the prevalence of keywords in titles} (e.g.~``editorial board'\,' or ``front cover''). Our method identifies a larger number of ``non-research'' articles than Unpaywall (12,284 versus 961), but of the 961 articles identified by Unpaywall, 99.8\% are also removed via our method.}. We thus decided to remove these articles as a source of uncertainty from our dataset for subsequent analyses.

Following the removal of articles without DEAL first and last authors, and articles with zero references or a single reference, our final dataset of articles from Dimensions was reduced to 410,084 articles (46\% of the original dataset).

For determing subject classifications, we used the ``Fields of Research'' (FOR) scheme available in Dimensions, which is itself based on the \href{https://dimensions.freshdesk.com/support/solutions/articles/23000018826-what-is-the-background-behind-the-fields-of-research-for-classification-system-}{Australian and New Zealand Standard Research Classification (ANZSRC) system}. The classification scheme consists of 22 divisions at the upper level. Unlike classification systems in other bibliometric databases which classify articles on a journal level, Dimensions firstly classifies articles on a single document level using a text-based classification approach. Where information is insufficient, Dimensions falls back to a journal-level classification. Some initial discussion of the strengths and weaknesses of the Dimensions approach has been conducted by \href{https://doi.org/10.1007/s11192-018-2855-y}{Bornmann (2018)}, and \href{https://doi.org/10.1007/s11192-018-2854-z}{Herzog and Kierkegaard Lunn (2018)}. \href{https://doi.org/10.1007/s11192-018-2855-y}{Bornmann (2018)} noted a number of inaccuracies in the classification of his own publication record in Dimensions. However, improvements to the classification system have since been implemented and Dimensions \href{https://www.dimensions.ai/release-notes/}{reported} an increase in the precision and recall of the method in August 2019.

Article records from Dimensions were matched to records in Crossref (for classification of Elsevier versus Non-Elsevier content, using the Crossref member ID of Elsevier, \href{https://www.crossref.org/members/prep/78}{78}) and Unpaywall (for determination of article OA status). Matching was conducted through exact matching of DOIs: 99.7\% and 99.8\% of articles in our dataset from Dimensions were subsequently matched to articles in Crossref and Unpaywall, respectively. Crossref data is based on an openly available Crossref database snapshot (\href{https://academictorrents.com/details/e4287cb7619999709f6e9db5c359dda17e93d515}{Crossref, 2021}) that contains all Crossref records registered until 7th January 2021. Relevant metadata fields were parsed applying the \emph{rcrossref} parsers (\href{https://CRAN.R-project.org/package=rcrossref}{Chamberlain et al., 2020}), following the same approach documented in \href{https://arxiv.org/abs/2102.04789}{Jahn et al.~(2021)}. To reduce computation time and storage demands, the Crossref dataset was subsequently limited to records registered after 1st January 2008. Unpaywall data is based on an openly available database snapshot (details available \href{https://unpaywall.org/products/snapshot}{here}) from February 2021. Processing of the Unpaywall dataset followed the same procedure as that documented in \href{https://edoc.hu-berlin.de/handle/18452/23336}{Hobert et al.~(2021)}.

\hypertarget{data-processing-storage-and-analysis}{%
\paragraph{Data Processing, Storage and Analysis}\label{data-processing-storage-and-analysis}}

To allow fast data processing and analysis, all large datasets described above (i.e.~those from Dimensions, Crossref and Unpaywall) were imported to \href{https://cloud.google.com/bigquery}{Google BigQuery}, a cloud data warehouse which allows querying of large datasets with SQL. All analysis of data was subsequently carried out in R (\href{http://www.R-project.org/}{R Core Team, 2020}), using the DBI (\href{https://CRAN.R-project.org/package=DBI}{R Special Interest Group on Databases et al., 2021}) and bigrquery (\href{https://CRAN.R-project.org/package=bigrquery}{Wickham and Bryan, 2020}) packages to interface Google BigQuery directly with R. Throughout this mostly automated data gathering and analysis process, tools from the Tidyverse (\href{https://doi.org/10.21105/joss.01686}{Wickham et al., 2019}) collection of packages for R were used.

\hypertarget{results}{%
\subsection{Results}\label{results}}

\hypertarget{publishing-behaviour-of-deal-researchers}{%
\subsubsection{Publishing behaviour of DEAL researchers}\label{publishing-behaviour-of-deal-researchers}}

In this section we assess how restricted access to Elsevier journals has influenced publishing patterns of researchers at DEAL institutions. Whilst access restrictions have reduced the ability for these researchers to read and download Elsevier articles, there exists no further barriers for them to \emph{publish} in Elsevier journals beyond those that previously existed, e.g.~meeting submission and peer-review criteria, affordability of journal-specific fees, etc. However, we hypothesise that access restrictions may lead to negative sentiment amongst researchers which would influence their decision when choosing a suitable publication venue for their work; such negative desire to publish with Elsevier was reported by 51\% of respondents of a survey conducted by the Bibsam Consortium when Elsevier restricted access to their journals in Sweden (\href{http://doi.org/10.1629/uksg.507}{Olson et al., 2020}).

We assess changes in publishing behaviour of DEAL researchers primarily through two related metrics: (1) the total number of articles published in Elsevier versus non-Elsevier journals each year, and (2) the year-on-year (YOY) change in the absolute proportion of articles published in Elsevier journals (i.e.~the change in Elsevier's market share of DEAL articles)\footnote{As an example, if DEAL researchers published 1000 articles in Year One, of which 200 were published in Elsevier journals, and 1500 articles in Year Two, of which 240 were published in Elsevier journals, then the change in proportion from Year One to Year Two is calculated as (240/1500) - (200/1000), equal to -0.04 and interpreted as a year-on-yearmarket share loss of 4\%.}. With respect to both of these metrics, we consider variation with respect to the year of contract cancellation of individual DEAL institutions, research disciplines, collaboration patterns and article OA status. We focus on the period between 2012 and 2020, allowing us to capture the long-term trends in the years prior to access restrictions, and the effect of access restrictions in 2018 on publishing patterns in two subsequent years.

\begin{figure}

{\centering \includegraphics[width=0.75\linewidth]{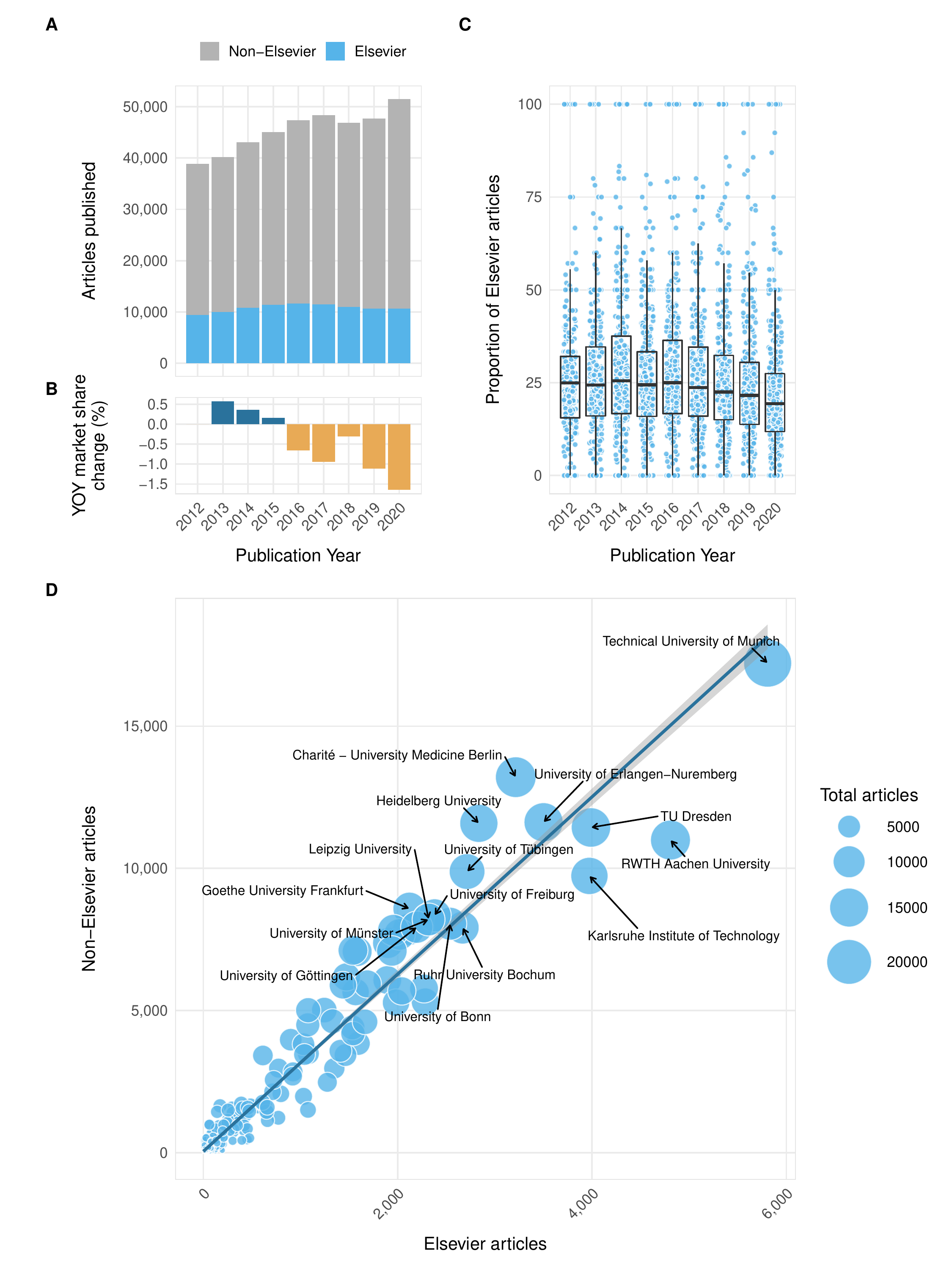} 

}

\caption{Publishing behaviour of DEAL researchers, 2012-2020. (A) Total number of articles published by DEAL researchers in Elsevier and non-Elsevier journals. (B) Year-on-year (YOY) change in Elsevier's market share of articles published by DEAL researchers. (C) Proportion of articles published by individual DEAL institutions in Elsevier journals. Boxplot horizontal lines denote lower quartile, median, upper quartile, with whiskers extending to 1.5*IQR. Points denote individual institutions, with added horizontal jitter for visibility. (D) Number of Elsevier versus Non-Elsevier articles published for each individual DEAL institution (totals aggregated for years 2012-2020). Point size is scaled by the total number of articles published by an institution. Institutions that published >10,000 articles in total are labelled.}\label{fig:items-publisher-year}
\end{figure}

Figure \ref{fig:items-publisher-year}A shows the change in the number of articles published by DEAL researchers between 2012 and 2020 (limited to articles with DEAL first and last authors, and with \textgreater1 reference). The total number of articles published per year increased during this period, from 38,849 articles in 2012 to 51,510 articles in 2020. In comparison, the number of
articles published in Elsevier journals increased from 9,401 articles in 2012 to 11,651 articles in 2016, and subsequently decreased to 10,623 articles in 2020. In terms of Elsevier's market share of DEAL articles (Figure \ref{fig:items-publisher-year}B), the years 2013-2015 show a trend of relatively small YOY market share gains (\textless0.5\% per year), with Elsevier's market share reaching a peak of 25.3\% in 2015. Subsequent years were characterised by a trend of larger market share losses, resulting in a final market share of 20.6\% in 2020. Whilst the general trend appears to occur independently of the timing of Elsevier access restrictions in mid-2018, the largest YOY losses occurred in 2019 (--1.1\%) and 2020 (--1.6\%), in the two years following the implementation of access restrictions. It is important to note that these results reflect the numbers and proportions of articles \emph{published} in journals, but do not necessarily reflect article \emph{submission} dynamics: articles take many months to proceed through peer-review and publication processes (and these processes are generally faster in STM fields versus social sciences/arts/humanities/economics fields; \href{https://doi.org/10.1016/j.joi.2013.09.001}{Björk and Solomon, 2013}), and acceptance/rejection rates may not have remained static or proportional over time.

Figure \ref{fig:items-publisher-year}C shows changes in the proportion of articles from each individual DEAL institution that are published in Elsevier journals. Patterns broadly reflect those shown in Figures \ref{fig:items-publisher-year}A and \ref{fig:items-publisher-year}B, with the proportion of articles in Elsevier journals remaining relatively static from 2012-2016, and subsequently declining. Interestingly, a small number of DEAL institutions appear to publish 100\% of their articles in Elsevier journals - upon inspection we find that these reflect institutions with extremely low publication volumes (\textless10 articles in a given year). Figure \ref{fig:items-publisher-year}D shows the total number of Elsevier versus non-Elsevier articles published by DEAL institutions aggregated over the entire time period 2012-2020. Results show general consistency in the proportion of articles published in Elsevier journals between large and small institutions; however, institutions of similar sizes also show sizeable variation. For example, Charité - University Medicine Berlin and RWTH Aachen University both published on the order of \textasciitilde16,000 articles between 2012 and 2020, yet only \textasciitilde24\% of articles published by Charité were published in Elsevier journals compared to \textasciitilde44\% by RWTH. Such large differences may reflect different research focuses of individual institutions, e.g.~Charité has a strong biomedical focus, whilst RWTH is a technical university with a historically strong focus in natural sciences, technology and engineering.

\hypertarget{did-publishing-patterns-differ-at-institutions-whose-contracts-with-elsevier-expired-in-different-years}{%
\paragraph{Did publishing patterns differ at institutions whose contracts with Elsevier expired in different years?}\label{did-publishing-patterns-differ-at-institutions-whose-contracts-with-elsevier-expired-in-different-years}}

\begin{figure}

{\centering \includegraphics[width=0.75\linewidth]{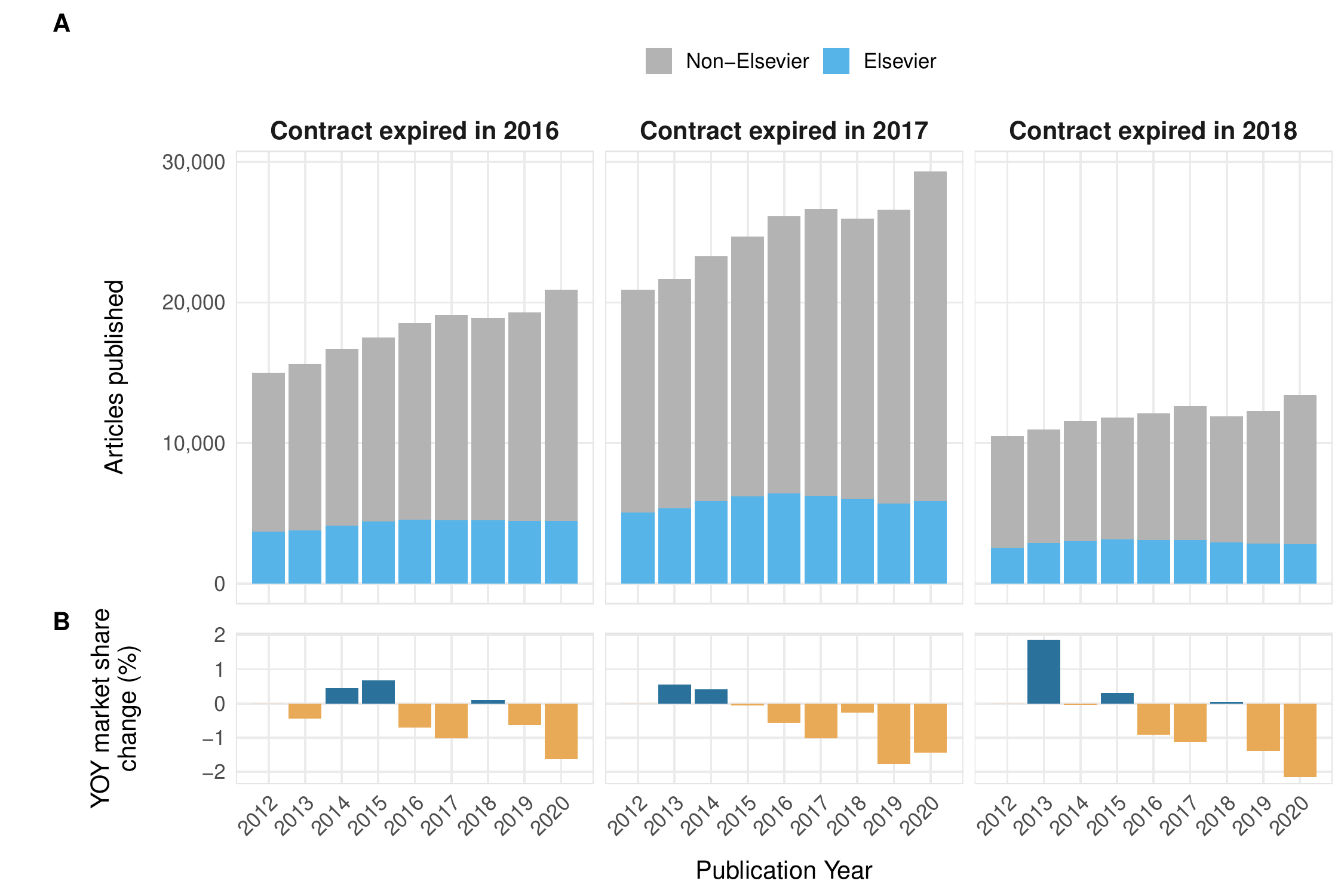} 

}

\caption{Changes in publishing behaviour of DEAL researchers, 2012-2020, dependent on year of contract expiration with Elsevier. (A) Total number of articles published by DEAL researchers in Elsevier and non-Elsevier journals. (B) year-on-year (YOY) change in Elsevier's market share of articles published by DEAL researchers.}\label{fig:items-publisher-year-cancellation}
\end{figure}

Our dataset contains three groups of DEAL institutions whose contracts with Elsevier expired at different time points: one group at the end of 2016 (N = 74), one at the end of 2017 (N = 110) and one at the end of 2018 (N = 204). In Figure \ref{fig:items-publisher-year-cancellation} we explore differences in publishing patterns between these different groups, to examine whether those whose contracts expired earlier showed different effects to those whose contracts expired more recently. All three groups grew in total publication volume between 2012 and 2020; the largest group by publication volume was those whose contracts expired in 2017, and the smallest those whose contracts expired in 2018. With respect to the share of articles published in Elsevier journals, all three groups display similar dynamics with an overall gain in Elsevier's market share between 2012 and 2015, and an overall loss between 2016 and 2020, although the exact magnitude and patterns of YOY gains and losses differs between each group. Between 2015 and 2017 (i.e.~two years prior to the access restrictions in 2018), Elsevier's market share changed by -1.7\%, -1.6\% and -2\% for the group whose contracts expired in 2016, 2017 and 2018, respectively; the rate of market share losses increased for all three groups between 2018 and 2020 (i.e.~two years following the access restrictions in 2018) to -2.3\%, -3.2\% and -3.5\%. A reason for the relatively homogenous behaviour may be that although the contracts expired at different timepoints, the time at which access to Elsevier was restricted was relatively similar across all DEAL institutions; those whose contracts expired at the end of 2016 and 2017 lost access in July 2018 (not including a brief 6-week period at the beginning of 2017), whilst those whose contracts expired at the end of 2018 lost access from the beginning of 2019 onwards.

\hypertarget{how-did-publishing-behaviour-vary-with-respect-to-research-disciplines}{%
\paragraph{How did publishing behaviour vary with respect to research disciplines?}\label{how-did-publishing-behaviour-vary-with-respect-to-research-disciplines}}

A complicating factor in our dataset is that we have included articles covering multiple research disciplines; we have already shown in Figure \ref{fig:items-publisher-year}D that variation in publishing patterns exists on the institutional level, which may be a consequence of the individual institutions' research focuses. A recent analysis of the effect of the agreements made between DEAL and Springer Nature, and DEAL and Wiley (\href{https://www.cesifo.org/en/publikationen/2021/working-paper/impact-german-deal-competition-academic-publishing-market}{Haucap et al., 2021}), chose to focus on a single discipline, Chemistry, with the justification that:

\begin{quote}
\emph{``Manuscript turnaround times differ substantially between different fields of science and are rather long in some disciplines such as economics (see, e.g, Ellison, 2002). Hence, the vast majority of articles published in economics journals in 2019 and 2020 will have been submitted before the DEAL agreements were announced. Therefore, our analysis focuses on the field of chemistry which has much faster turnaround times so that we can expect the DEAL agreements to already have at least some impact.''}
\end{quote}

\begin{figure}

{\centering \includegraphics{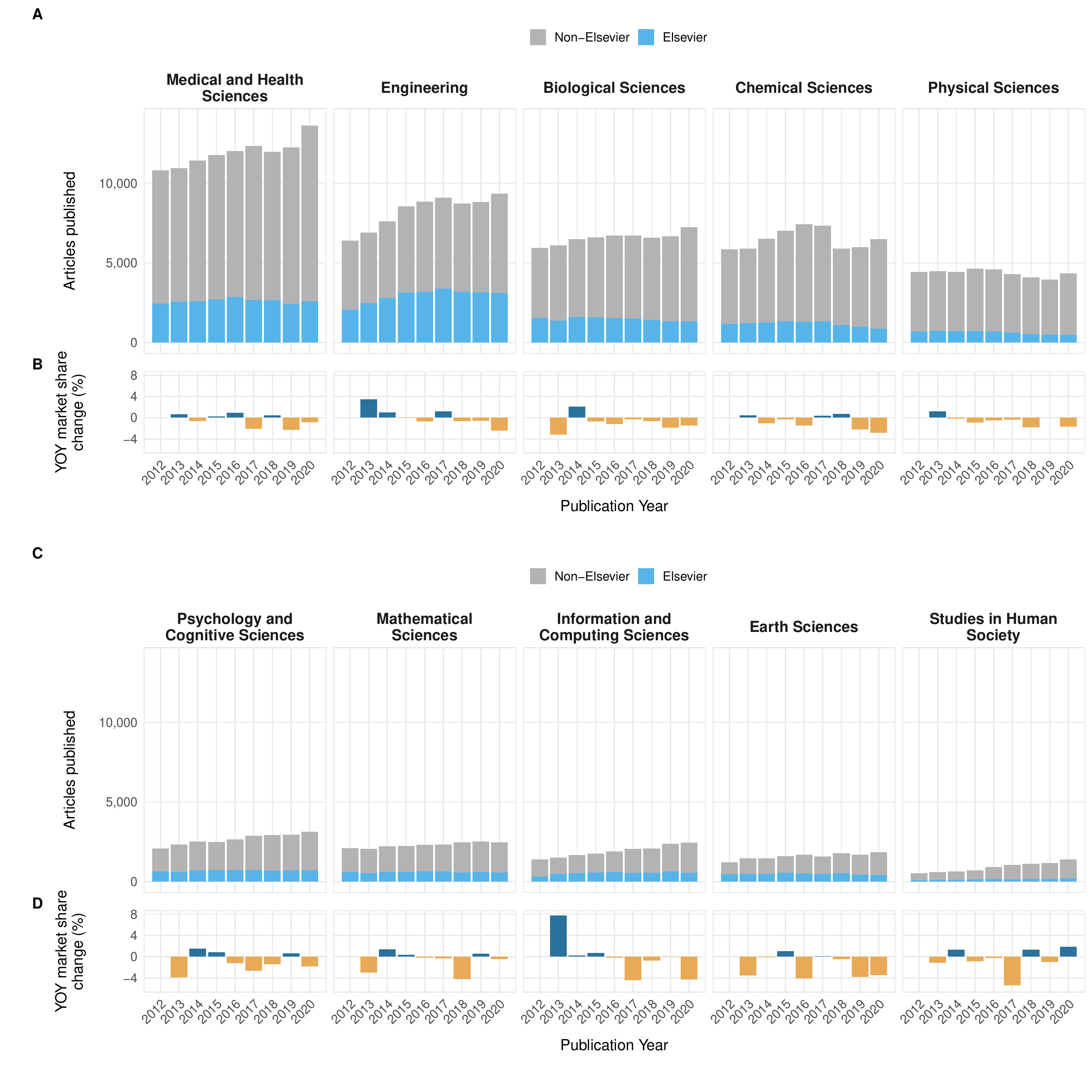} 

}

\caption{Changes in publishing behaviour of DEAL researchers, 2012-2020, dependent on research discipline (top-10 by publishing volume are shown). (A and C) Total number of articles published by DEAL researchers in Elsevier and non-Elsevier journals for individual disciplines. (B and D) year-on-year (YOY) change in Elsevier's market share of articles published by DEAL researchers.}\label{fig:items-publisher-year-category}
\end{figure}

Our analysis covers a longer time period than that of \href{https://www.cesifo.org/en/publikationen/2021/working-paper/impact-german-deal-competition-academic-publishing-market}{Haucap et al.~(2021)}, and we assess effects covering the entire time period in which negotiations with Elsevier began in 2016, the time at which access was restricted in July 2018, and two subsequent publication years thereafter until the end of 2020. We therefore feel justified in including a broader range of disciplines in our approach with longer publishing timelines than Chemistry. Nonetheless, we have also analysed changes at the level of individual disciplines (i.e.~Dimensions Fields of Research) (Figure \ref{fig:items-publisher-year-category}). For visualisation purposes we limited results here to the top-10 disciplines by publishing volume, which tend to focus primarily on STM disciplines; full results for all 22 disciplines can be found in the supplementary material at archived on Zenodo: \url{https://doi.org/10.5281/zenodo.4771576}.

Overall, patterns of publishing behaviour for individual research disciplines display a higher degree of fluctuation than our overall analysis with all disciplines aggregated (likely due to smaller sample sizes). Most disciplines show overall growth in the total number of articles published over the timeframe of our analysis, although publication volumes in some disciplines appear to have decreased or reached a plateau in recent years (e.g.~Chemical Sciences, Physical Sciences). The results also reveal variation in the tendency of authors to publish in Elsevier journals between disciplines: aggregated over the entire period between 2012 and 2020, the discipline with the highest proportion of articles published in Elsevier journals is Economics (41.9\%), and the lowest proportion is in Philosophy and Religious Studies (5.5\%). In 18 of the 22 disciplines, Elsevier lost overall market share between 2012 and 2020; the largest losses are reported in Earth Sciences (--14.4\%), Language, Communication and Culture (--13\%), History and Archaeology (--11.4\%), and Agricultural and Veterinary Sciences (--10.3\%); conversely, four individual disciplines showed market share gains over the same time period (Engineering, +1.1\%; Economics, +2.8\%; Commerce, Management, Tourism and Services, +4.5\%; Technology, +8.2\%). However, the patterns of YOY growth/losses in Elsevier's market share for individual disciplines varies significantly and we do not observe any consistent trends that can be clearly attributed to the access restrictions in 2018; only 10 of the 22 disciplines showed greater market share losses in the two years following the access restrictions compared to the two years prior to access restrictions.

\hypertarget{how-did-publishing-behaviour-vary-with-respect-to-collaboration-patterns}{%
\paragraph{How did publishing behaviour vary with respect to collaboration patterns?}\label{how-did-publishing-behaviour-vary-with-respect-to-collaboration-patterns}}

\begin{figure}

{\centering \includegraphics{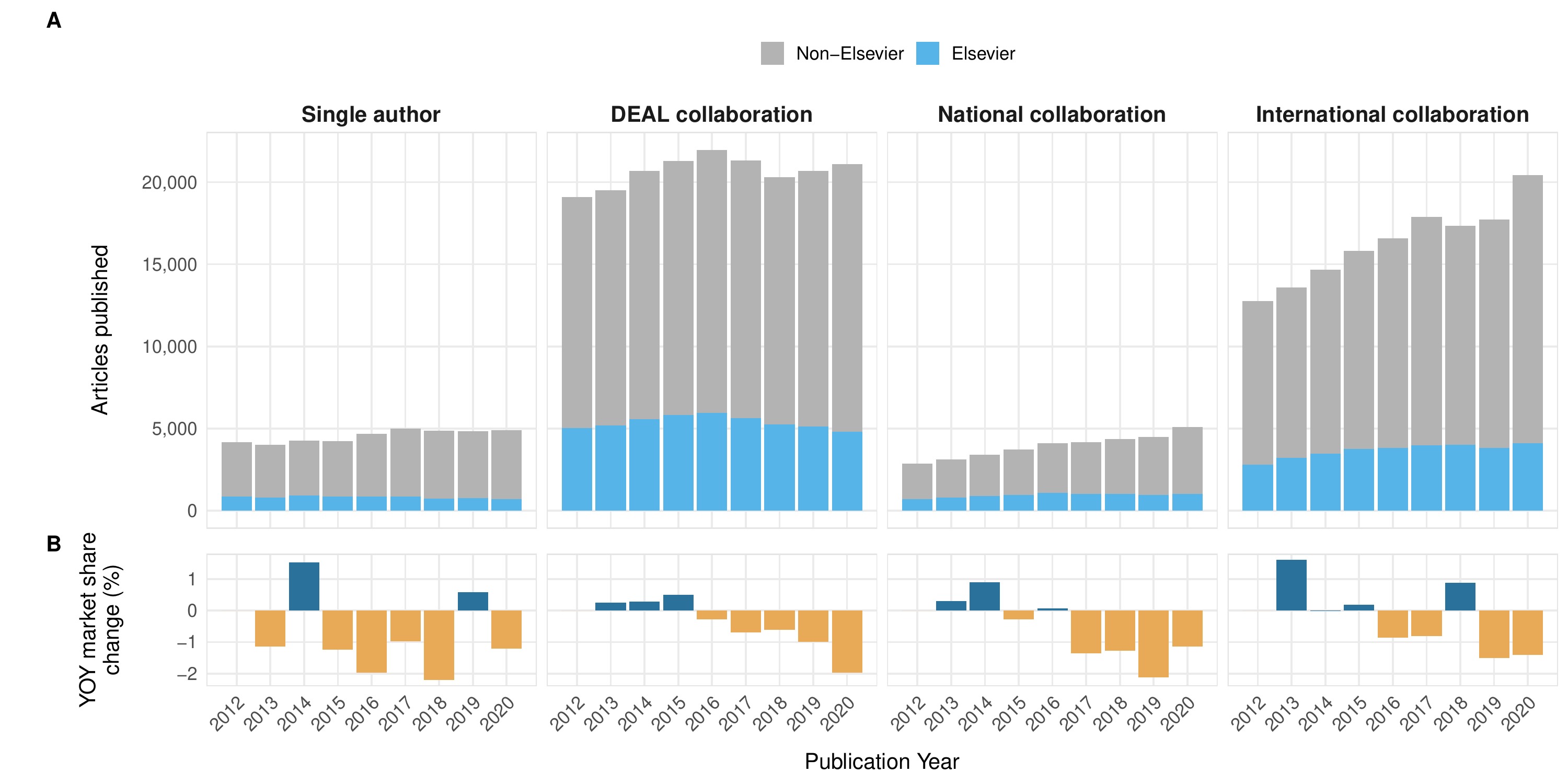} 

}

\caption{Changes in publishing behaviour of DEAL researchers, 2012-2020, dependent on collaboration status. ``Single author'' refers to articles that are authored by a single researcher, ``DEAL collaboration'' refers to articles with multiple authors, where all authors are based exclusively at DEAL institutions, ``National collaboration'' refers to articles where some authors are based at DEAL institutions and others at non-DEAL institutions within Germany, and ``International collaboration'' refers to articles where some authors are based at DEAL institutions and others at institutions outside of Germany. (A) Total number of articles published by DEAL researchers in Elsevier and non-Elsevier journals. (B) year-on-year (YOY) change in Elsevier's market share of articles published by DEAL researchers.}\label{fig:items-publisher-year-collaboration}
\end{figure}

Another potential confounder of our overall results in earlier sections relates to that of collaboration behaviour. An article written solely by researchers at DEAL institutions may have a different publication strategy compared to one written in an internationally-collaborative project, where international colleagues may be less knowledgeable of Elsevier access restrictions, or less disturbed in their daily research activities. Figure \ref{fig:items-publisher-year-collaboration} shows changes in publishing behaviour with respect to collaboration status. We classified articles into four distinct collaboration classes: (1) ``Single author'', referring to articles published by only a single researcher, (2) ``DEAL collaboration'', referring to articles with multiple authors where all authors of the article are based exclusively at DEAL institutions, (3) ``National collaboration'', referring to articles where some authors are based at DEAL institutions and others at non-DEAL institutions within Germany, and (4) ``International collaboration'', referring to articles where some authors are based at DEAL institutions and others at institutions outside of Germany. Note that for all of these classes, the first and last authors (or the single author) always have an affiliation at a DEAL institution. For all of the collaborative classes (i.e.~ignoring the single-author class) the minimum number of authors per article was 2, the median number of authors was 3, 6 and 6 for DEAL collaborations, national collaborations and international collaborations, respectively, and the maximum number of authors was 68, 48 and 729 for DEAL collaborations, national collaborations and international collaborations, respectively.

Overall, the most publications in our datasets are produced in DEAL collaborations or international collaborations, whilst the number of articles published by single-authors and in other national collaborations is comparatively small. The number of single-author articles has remained relatively static over time, only increasing from 4,164 in 2012 to 4,898 in 2020. With respect to DEAL-only collaborations, the total number of published articles grew from 19,099 in 2012 to 21,948 in 2016, but subsequently plateaued or even slightly declined, with 21,098 articles published in 2020. In comparison, the number of articles published as national collaborations and international collaborations grew substantially from 2012 to 2020 (from 2,842 to 5,080 for national collaborations, and 12,744 to 20,434 for international collaborations). The findings suggest that over time, DEAL researchers are transitioning towards a more collaborative research environment, particularly with respect to increasing collaborations with international partners, which echoes similar trends observed across Europe (\href{https://doi.org/10.1080/03075079.2020.1749254}{Kwiek, 2020}).

In terms of Elsevier's market share, single-author articles show a general long-term trend towards a market share loss over time (from an overall market share of 20.8\% in 2012 to 14.2\% in 2020), albeit punctuated by two YOY market share gains in 2014 and 2019. All three collaborative groups also display overall losses between 2012 and 2020 (from an overall market share of 26.3\%, 24.8\%, and 22\% in 2012, to 22.8\%, 19.9\%, and 20.1\% in 2020 for DEAL collaborations, national collaborations and international collaborations, respectively), yet the patterns and timing differ somewhat between different collaboration groups. The share of articles published as DEAL collaborations decreased relatively steadily from 2016 to 2019, and then much more sharply in 2020 (YOY market share loss of -2\%). In comparison, the market share of articles published as national collaborations only began to decrease in 2017, and the overall market share decrease for articles published as international collaborations between 2016 and 2020 was punctuated by a YOY increase of 0.9\% in 2018.

\hypertarget{how-did-publishing-behaviour-vary-with-respect-to-oa-status}{%
\paragraph{How did publishing behaviour vary with respect to OA status?}\label{how-did-publishing-behaviour-vary-with-respect-to-oa-status}}

Between 2010 and 2018 the proportion of articles authored by all researchers at German institutions that were made OA increased dramatically, from 27\% in 2010 to 52\% in 2018 (\href{https://doi.org/10.1007/s11192-021-04002-0}{Hobert et al., 2021}). We also aimed to determine whether the restriction of access to Elsevier journals had an effect on OA publishing behaviour of DEAL researchers in Elsevier journals, with the hypothesis that increased awareness of access issues, and motivation to ensure accessibility for colleagues, would motivate DEAL researchers to publish articles under OA licenses. Articles were classified into OA categories following the same {[}schema used by Unpaywall{]}((\url{https://support.unpaywall.org/support/solutions/articles/44001777288-what-do-the-types-of-oa-status-green-gold-hybrid-and-bronze-mean-}): broadly, ``Gold'' refers to articles published in fully OA journals, ``Hybrid'' to articles published under a OA license in an otherwise subscription-based journal, ``Green'' to articles that have been made available in an OA repository, and ``Bronze'' to articles that are freely accessible on the publisher's website but are not published under an OA license. All articles that are not freely accessible are classified as ``Closed''. An important point for the analysis of OA shares is that our dataset measures OA availability at the time of measurement (in our case, February 2021), and so OA shares do not necessarily reflect the OA status of an article at the time of its publication (e.g.~an article could transition from Closed to Green several years after publication, if a version is deposited to an OA repository after an embargo period).

\begin{figure}

{\centering \includegraphics{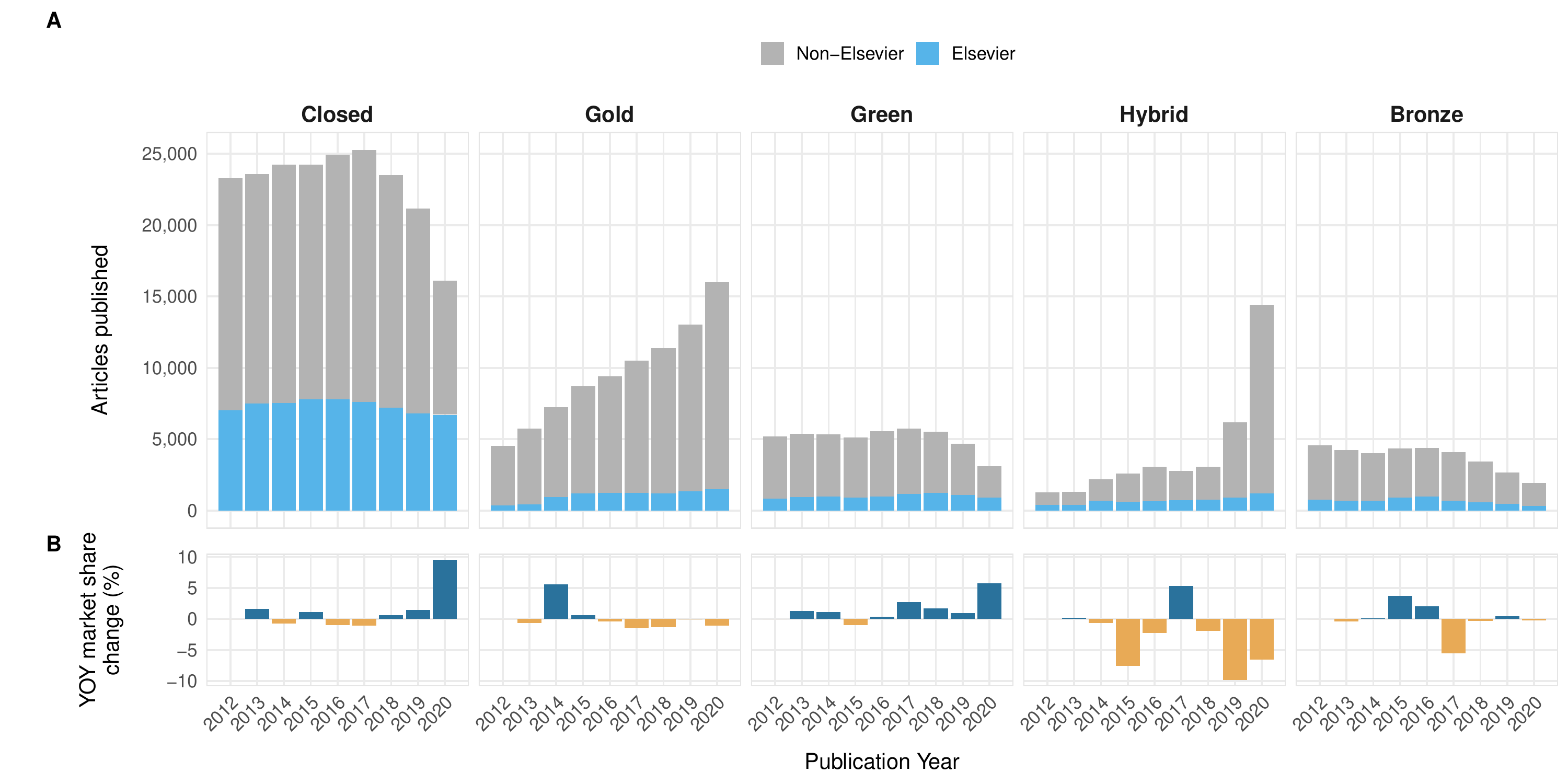} 

}

\caption{Changes in publishing behaviour of DEAL researchers, 2012-2020, dependent on OA status. (A) Total number of articles published by DEAL researchers in Elsevier and non-Elsevier journals. (B) year-on-year (YOY) change in Elsevier's market share of articles published by DEAL researchers.}\label{fig:items-publisher-year-oa}
\end{figure}

OA publishing patterns of DEAL institutions shown in Figure \ref{fig:items-publisher-year-oa} broadly agree with the previous findings of \href{https://doi.org/10.1007/s11192-021-04002-0}{Hobert et al.~(2021)} which focused on a larger number of German universities and non-university research institutions. We find that the total proportion of OA articles (including all OA categories analysed) published by DEAL researchers grew from 40\% in 2012 to 68.8\% in 2020, compared to a growth in OA articles of 27\% in 2010 to 52\% in 2018 reported by \href{https://doi.org/10.1007/s11192-021-04002-0}{Hobert et al.~(2021)}. The strong growth of OA at DEAL institutions is driven largely by the growth of Gold OA from 2012 (4,544 articles) to 2020 (15,986 articles), and of hybrid OA in particular from 2018 (3,067 articles) to 2020 (14,376 articles), which may be driven, at least in part, by new agreements made with publishers including Springer Nature and Wiley (\href{https://www.cesifo.org/en/publikationen/2021/working-paper/impact-german-deal-competition-academic-publishing-market}{Haucap et al., 2021}).

With respect to publishing patterns in Elsevier journals, the growth of OA has been relatively moderate compared to the overall picture in Germany: from 2012 to 2020 the total proportion of DEAL articles published in Elsevier journals that were made OA increased only from 25.2\% to 36.8\%. Regarding individual OA categories, we observe the largest changes occurring from around 2014 onwards in Elsevier journals: the number of Gold articles increased from 950 articles in 2014 to 1,497 articles in 2020, and the number of Hybrid articles increased from 681 articles in 2014 to 1,188 articles in 2020. In terms of market share, the most prominent feature is that of YOY losses for Elsevier in the Hybrid OA market share of \textgreater5\% in consecutive years in 2019 and 2020, however, this appears to be driven more by the surge of Hybrid OA publishing in other venues rather than a reduction in the volume of Hybrid OA published by Elsevier (which is also reflected in the large gain of market share (+9.5\%) for Elsevier of closed articles in 2020). Another interesting feature is the growth of Green OA in Elsevier journals in 2020 (+5.7\% market share), suggesting an increasing proportion of researchers publishing in Elsevier journals are depositing their work to OA repositories in comparison to across all publishers as a whole.

\hypertarget{citing-behaviour-of-deal-researchers}{%
\subsubsection{Citing behaviour of DEAL researchers}\label{citing-behaviour-of-deal-researchers}}

A number of studies have found a citation advantage of open-access publications over their closed-access counterparts (c.f. \href{https://doi.org/10.7717/peerj.4375}{Piwowar et al., 2018}), with the implication that having access to articles makes them easier to read, download and ultimately more likely to be cited. If this were true, we would expect that restricting access to a set of articles would have the opposite effect, i.e.~reduce their ability to be cited. We therefore aimed to investigate the effect of Elsevier access restrictions on DEAL researchers citing behaviour, using a set of 16,919,143 references (DOI-DOI links) from articles authored by DEAL researchers (limited to articles where the first and last author was at a DEAL institution, and articles that contained more than a single reference: see previous ``Methods'' section for details).

\begin{figure}

{\centering \includegraphics[width=0.5\linewidth]{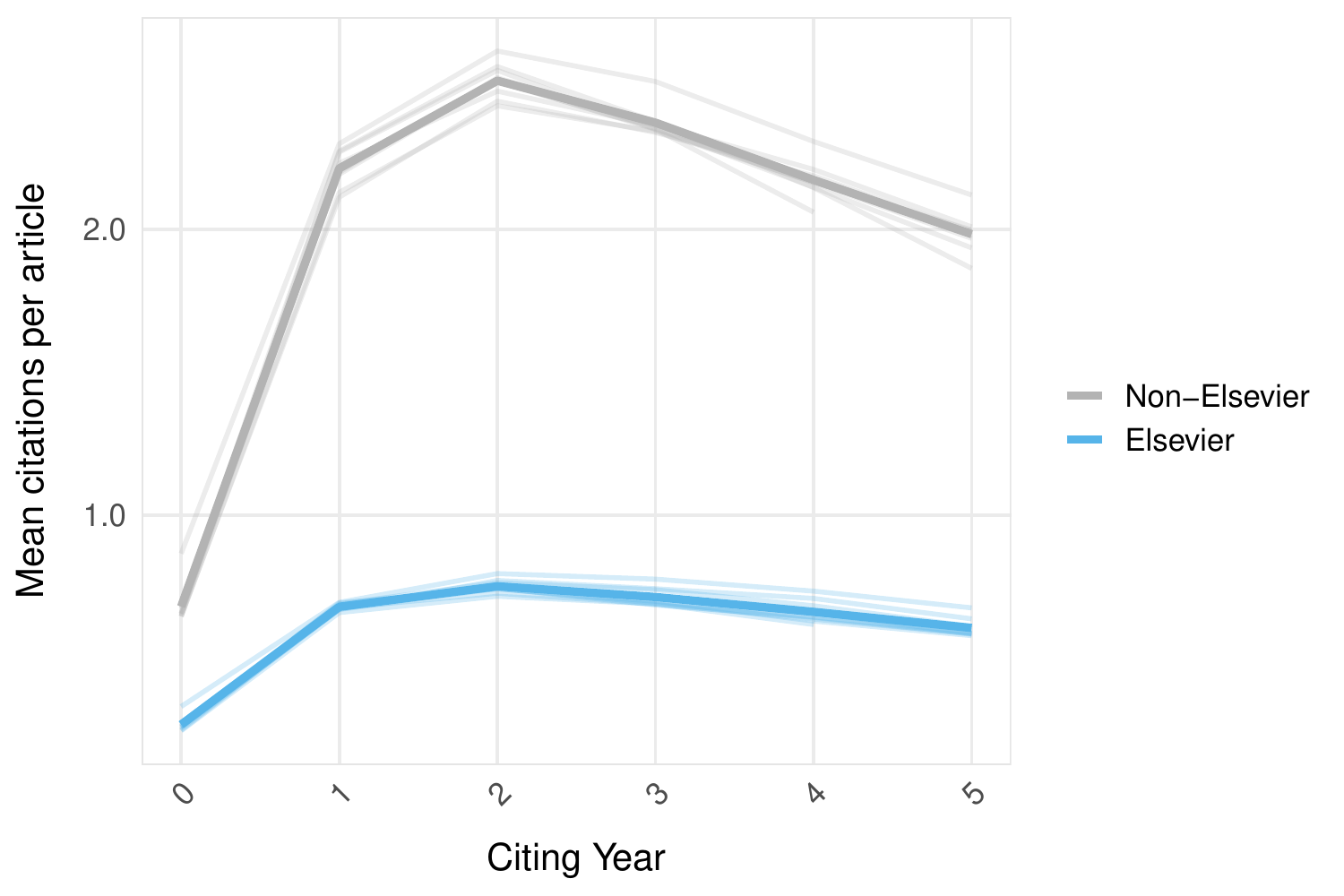} 

}

\caption{Citing dynamics for articles published by DEAL researchers. Lines show the evolution of the mean number of citations made to Elsevier or non-Elsevier articles, as a function of citing year. Thick lines represent the overall mean, thin lines represent articles from individual publication years.}\label{fig:references-publisher-year-cityear}
\end{figure}

A complicating factor in the analysis of citing behaviour is that there exists time variation in both the year of publication of an article, and in the publication year of the articles they cite, i.e.~articles may cite other articles published in any prior year. An overview of citing dynamics for our dataset of DEAL articles is displayed in Figure \ref{fig:references-publisher-year-cityear}, where we show the mean number of citations per article made to Elsevier or non-Elsevier articles, as a function of citing year (where citing year refers to the difference in years between the publication date of the citing article and the publication date of the cited article). An important point for this section is that we study the problem in an inverse way to the majority of citation studies: we analyse \emph{outgoing} rather than \emph{incoming} citations. Nonetheless our results display similarly typical dynamics (see, e.g.~\href{https://doi.org/10.1016/j.joi.2015.07.006}{Parolo et al.~(2015)}): articles in our dataset cite relatively few articles published in the same year (presumably, as citing articles must first be written and proceed through the lengthy peer-review process), the number of cited articles peaks in citing years 2-3, and subsequently slowly declines over the following years.

For the purposes of the following analyses, we make the assumption that citing behaviour in response to Elsevier access restrictions is most likely to change for recent citations, which we define as citations with a citation age less than or equal to 2 years. This assumption is based on the fact that access restrictions affected new journal issues at all DEAL institutions, whilst only a subset of DEAL institutions also lost access to their back-catalogue of articles (\href{https://doi.org/10.1126/science.355.6320.17}{Vogel, 2017a}). We may therefore reasonably expect that authors who have had access restricted to Elsevier journals post-2018 either still have access to the older back-catalogues, or may have saved older articles to local storage media (e.g.~in reference management software) during the time when access was still available. For assessing changes in citing behaviour, we define three related metrics of measurement in similar way to our analysis of publishing behaviour: (1) the proportion of articles published by DEAL researchers that cited \emph{any} article in an Elsevier journal, (2) the total number of citations made to articles in Elsevier versus non-Elsevier journals from articles published by DEAL researchers, and (3) the annual change in the absolute proportion of citations made to articles in Elsevier journals from articles published by DEAL researchers (which, for the purposes of this analysis, we will term as Elsevier's ``market share'' of citations).

\begin{figure}

{\centering \includegraphics[width=0.833\linewidth]{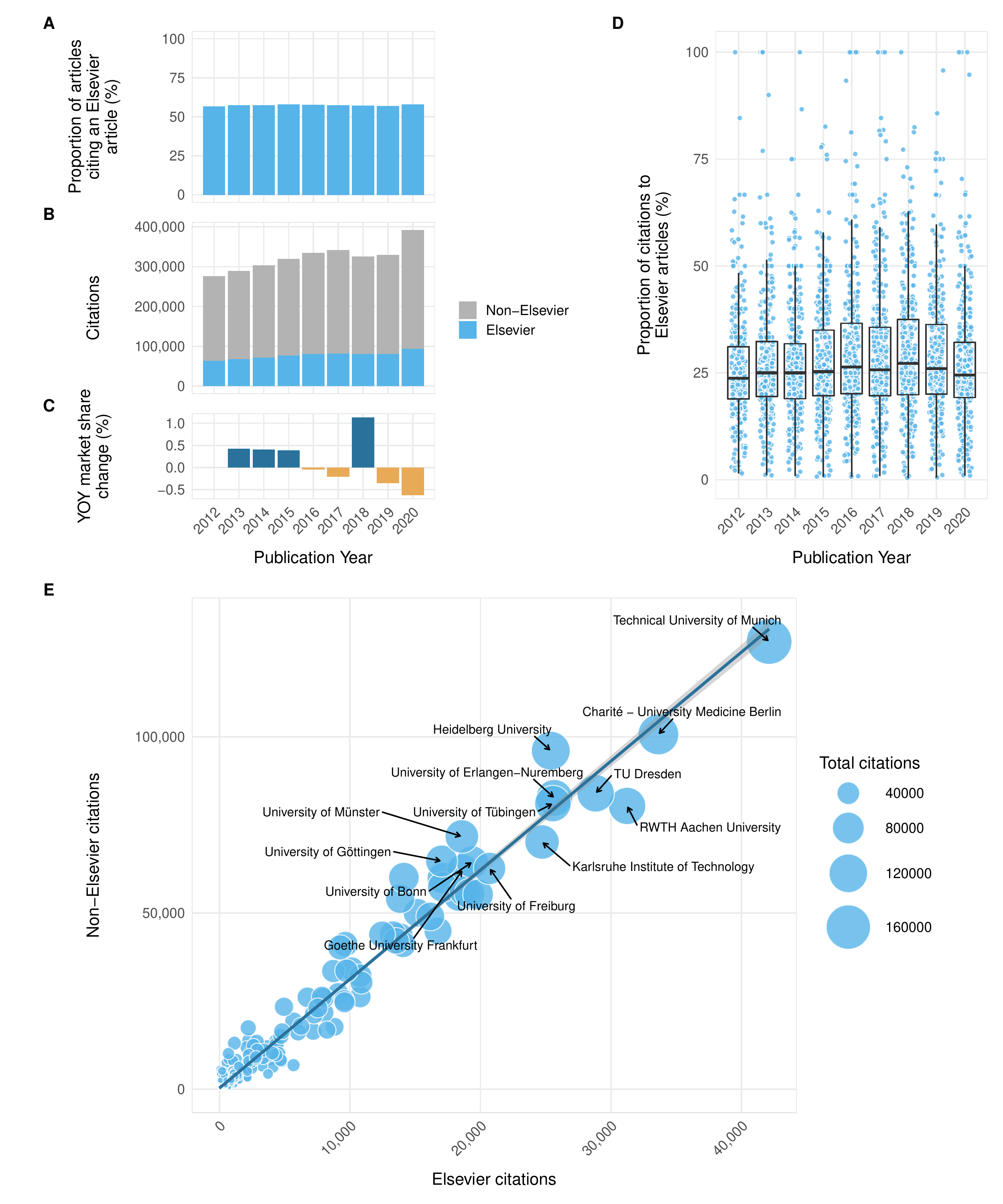} 

}

\caption{Citing behaviour of DEAL researchers, 2012-2020. All data is based on citations from articles published by DEAL researchers to articles with a citation age of 2 years or less. (A) Proportion of articles published by DEAL researchers that cite at least one single Elsevier article. (B) Total number of citations from articles published by DEAL researchers to articles published in Elsevier versus non-Elsevier journals. (C) Year-on-year (YOY) change in Elsevier's market share of citations from articles published by DEAL researchers. (D) Proportion of citations from articles published by individual DEAL institutions to articles in Elsevier journals. Boxplot horizontal lines denote lower quartile, median, upper quartile, with whiskers extending to 1.5*IQR. Points denote individual institutions, with added horizontal jitter for visibility. (E) Number of citations to articles in Elsevier versus Non-Elsevier journals for articles published by individual DEAL institutions (totals aggregated for years 2012-2020). Point size is scaled by the total number of citations from an institution. Institutions citing >80,000 articles are labelled.}\label{fig:references-publisher-year}
\end{figure}

Overall results of citing behaviour of DEAL researchers' between 2012 and 2020 are shown in Figure \ref{fig:references-publisher-year}, as well as citing behaviour for individual institutions. All results are based on a maximum citation age of two years. Figure \ref{fig:references-publisher-year}A shows the proportion of articles published in a given year, that cite at least a single Elsevier article. Overall, the proportion of articles that have cited an Elsevier article has remained relatively constant over time, fluctuating around a mean of 57.4\%. Figure \ref{fig:references-publisher-year}B shows that the total number of citations largely echoes trends of the number of articles published in the same year (Figure \ref{fig:items-publisher-year}A); however, whilst the total number of Elsevier articles published by DEAL researchers decreased from 2015 to 2020 (Figure \ref{fig:items-publisher-year}A), the number of citations made to Elsevier articles continued to rise, from 80,234 citations in 2016 to 93,840 citations in 2020, although this is largely driven by the large single-year peak in 2020. The result of these patterns is that over the entire time period of our analysis, Elsevier's market share of citations from DEAL researchers increased from 22.8\% in 2012 to 23.9\% in 2020, with a prominent gain in market share in 2018 (+1.1\%) and the largest market share loss occurring in 2020 (--0.6\%). The losses of market share in 2019 and 2020 are consistent with an expectation that reduced access to Elsevier articles post-2018 would have reduced the ability of researchers to cite those articles; however, in comparison to market share losses in publishing volume over the same time period (--2.7\%), the loss of market share of citations (--1\%) is relatively moderate.

Figures \ref{fig:references-publisher-year}D and \ref{fig:references-publisher-year}E display citing behaviour at the level of individual institutions. Overall, the proportion of citations made to articles in Elsevier journals reflect the same patterns of publishing behaviour as observed in Figures \ref{fig:items-publisher-year}C and \ref{fig:items-publisher-year}D - the proportion of citations made to articles in Elsevier journals gradually increased from 2012 to 2018, and small decreases were noted in 2019 and 2020. Aggregated over the entire 2012-2020 time period, we find little evidence of size-related effects: large research institutions generally cite articles in Elsevier journals in similar proportions to small research institutions (Figure \ref{fig:items-publisher-year}E), although some variation between individual institutions exists.

\hypertarget{did-citing-patterns-differ-at-institutions-whose-contracts-with-elsevier-expired-in-different-years}{%
\paragraph{Did citing patterns differ at institutions whose contracts with Elsevier expired in different years?}\label{did-citing-patterns-differ-at-institutions-whose-contracts-with-elsevier-expired-in-different-years}}

\begin{figure}

{\centering \includegraphics[width=0.833\linewidth]{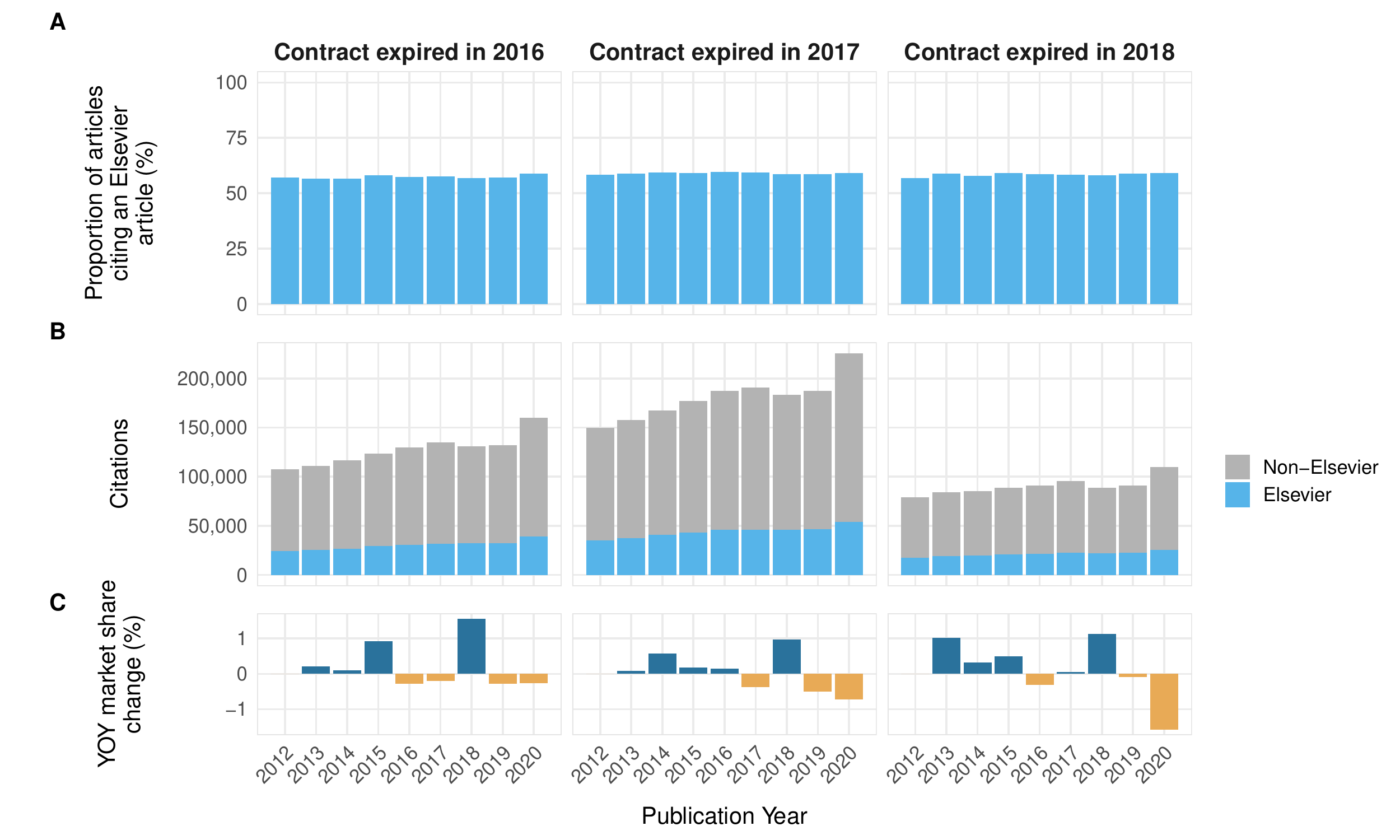} 

}

\caption{Citing behaviour of DEAL researchers, 2012-2020, dependent on year of contract expiration with Elsevier. All data is based on citations from articles published by DEAL researchers to articles with a citation age of 2 years or less. (A) Proportion of articles published by DEAL researchers that cite at least one single Elsevier article. (B) Total number of citations from articles published by DEAL researchers to articles published in Elsevier versus non-Elsevier journals. (C) Year-on-year (YOY) change in Elsevier's market share of citations from articles published by DEAL researchers.}\label{fig:references-publisher-year-cancellation}
\end{figure}

As with our analysis of publishing patterns (Figure \ref{fig:items-publisher-year-cancellation}), we also analysed citing behaviour of DEAL researchers dependent on the contract expiration date of their institution with Elsevier (Figure \ref{fig:references-publisher-year-cancellation}). Results show relatively homogenous behaviour between the different groups and reflect the overall results from Figure \ref{fig:items-publisher-year}. One prominent feature, however, is a YOY market share loss of -1.6\% in 2020 for the group whose contracts with Elsevier expired in 2018; in comparison, the groups whose contracts expired in 2016 and 2017 show only relatively small losses for the same year (-0.3\% and -0.7\%, respectively).

\hypertarget{how-did-citing-behaviour-vary-with-respect-to-research-disciplines}{%
\paragraph{How did citing behaviour vary with respect to research disciplines?}\label{how-did-citing-behaviour-vary-with-respect-to-research-disciplines}}

\begin{figure}
\centering
\includegraphics{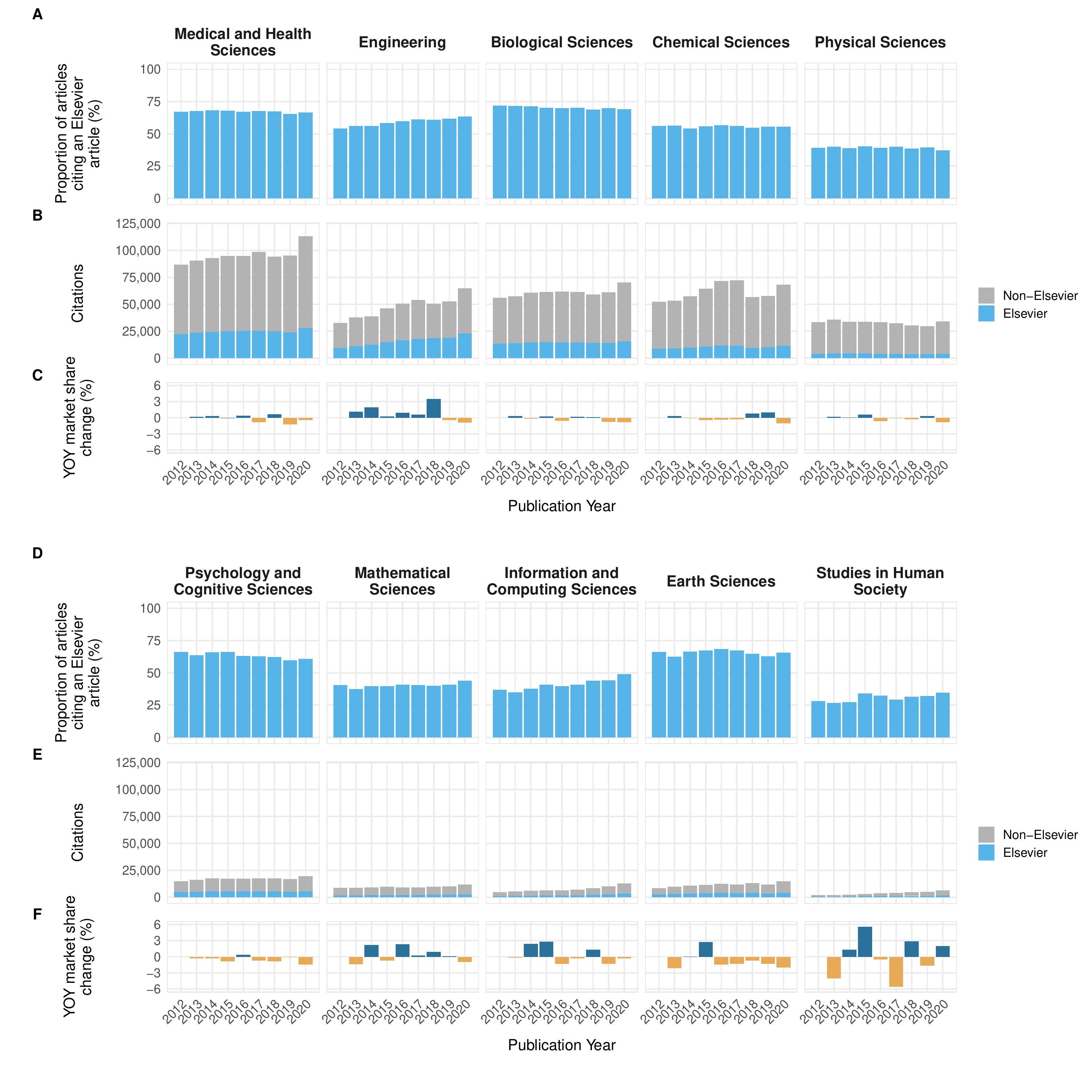}
\caption{\label{fig:references-publisher-year-category}Citing behaviour of DEAL researchers, 2012-2020, dependent on research discipline (top-10 disciplines by publication volume are shown). All data is based on citations from articles published by DEAL researchers to articles with a citation age of 2 years or less. (A and D) Proportion of articles published by DEAL researchers that cite at least one single Elsevier article. (B and E) Total number of citations from articles published DEAL researchers to articles published in Elsevier versus non-Elsevier journals. (C and F) Year-on-year (YOY) change in Elsevier's market share of citations from articles published by DEAL researchers.}
\end{figure}

We analysed how citing behaviour also varied with respect to individual research disciplines (Dimensions Fields of Research). Results for the top-10 research disciplines by publication volume are shown in Figure \ref{fig:references-publisher-year-category}, and full results for all 22 research disciplines are available in the supplementary material archived on Zenodo: \url{https://doi.org/10.5281/zenodo.4771576}.

We observe a high degree of variation between disciplines in their propensity to cite articles in Elsevier journals: aggregated across all publication years, the research disciplines with the highest proportion of citations to articles in Elsevier journals were Built Environment and Design (47.7\%) and Economics (43\%), whilst the disciplines that made the lowest proportion of citations to articles in Elsevier journals were Physical Sciences (12.1\%) and Philosophy and Religious Studies (11.6\%). We also analysed the proportion of articles that cited at least one Elsevier article for each individual discipline, and found the highest proportion in Environmental Sciences (74.3\%, aggregated across all publication years), and lowest proportion in Philosophy and Religious Studies (15.1\%). However, these results need to be carefully interpreted as the number of references per article (i.e.~the reference density) also varies between research disciplines (\href{https://doi.org/10.1016/j.joi.2017.11.003}{Sánchez-Gil et al., 2018}). Thus, articles in disciplines where the average number of references per article is higher are likelier to cite at least a single Elsevier article, all other factors being equal.

Given the high variability shown in YOY market shares for individual disciplines (Figure \ref{fig:references-publisher-year-category}C), it is difficult to interpret clear long-term trends that can be attributed directly to the Elsevier access restrictions in 2018. However, we do find that from 2018 to 2020, the share of citations made to articles in Elsevier journals decreased in 17 of the 22 disciplines (with the largest loss of -10.3\% in Built Environment and Design) - in only 5 disciplines did the share of references to articles in Elsevier journals increase during this time (Philosophy and Religious Studies, +1.5\%; Studies in Creative Arts and Writing, +0.8\%; Agricultural and Veterinary Sciences, +0.7\%; Studies in Human Society, +0.3\%; Commerce, Management, Tourism and Services, +0.2\%).

\hypertarget{how-did-citing-behaviour-vary-with-respect-to-collaboration-patterns}{%
\paragraph{How did citing behaviour vary with respect to collaboration patterns?}\label{how-did-citing-behaviour-vary-with-respect-to-collaboration-patterns}}

\begin{figure}

{\centering \includegraphics{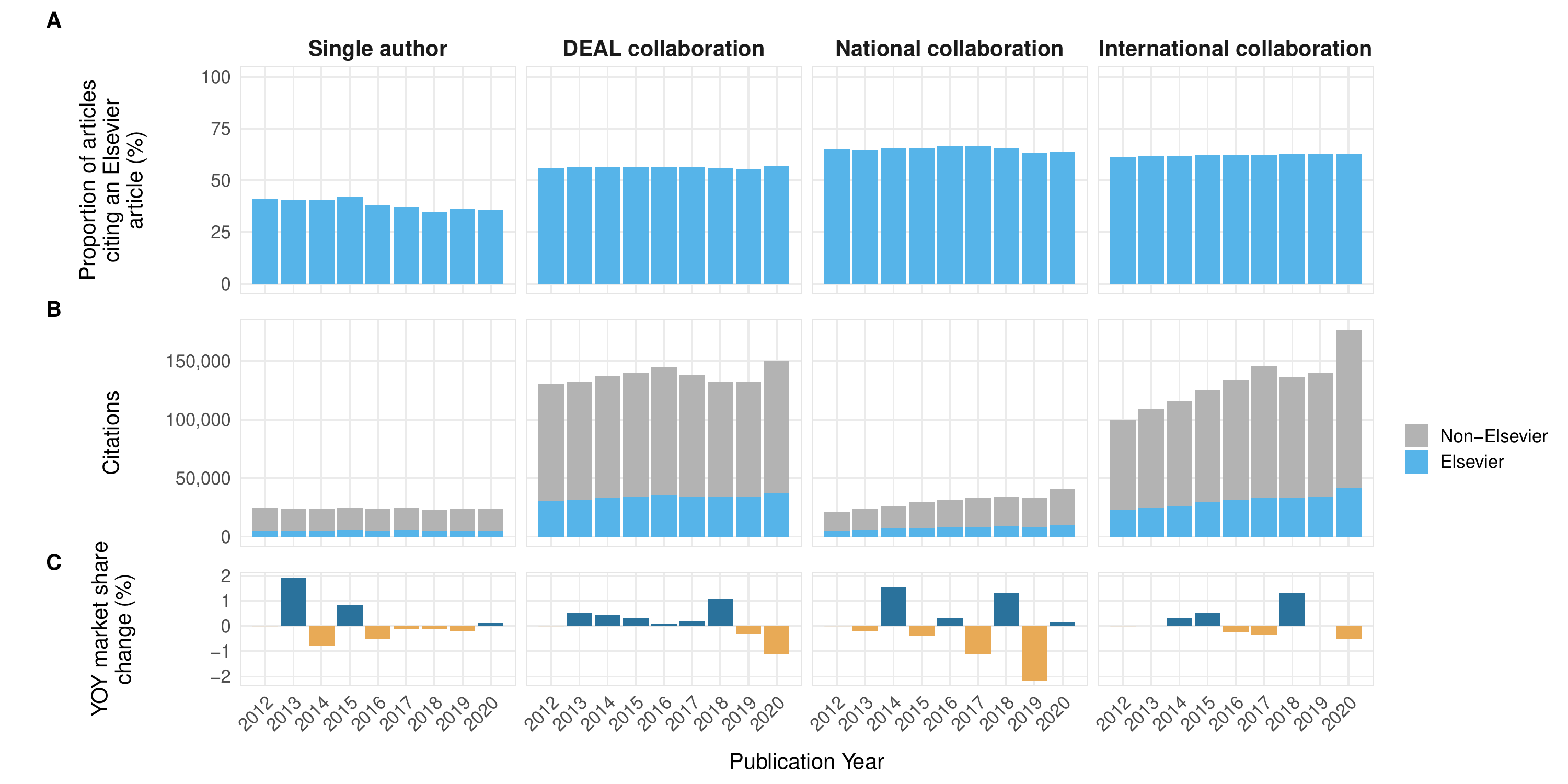} 

}

\caption{Citing behaviour of DEAL researchers, 2012-2020, dependent on collaboration status. ``Single author'' refers to articles that are authored by a single researcher, ``DEAL'' collaboration' refers to articles with multiple authors, where all authors are based exclusively at DEAL institutions, ``National collaboration'' refers to articles where some authors are based at DEAL institutions and others at non-DEAL institutions within Germany, and ``International collaboration'' refers to articles where some authors are based at DEAL institutions and others at institutions outside of Germany. All data is based on citations from articles published by DEAL researchers to articles with a citation age of 2 years or less. (A) Proportion of articles published by DEAL researchers that cite at least one single Elsevier article. (B) Total number of references in articles published DEAL researchers to articles published in Elsevier versus non-Elsevier journals. (C) Year-on-year (YOY) change in Elsevier's market share of references in articles published by DEAL researchers.}\label{fig:references-publisher-year-collaboration}
\end{figure}

Collaboration patterns may also have an influence on citing behaviour. Some evidence exists (e.g.~\href{https://doi.org/10.1371/journal.pone.0049176}{Milojević, 2012}) that authors who collaborate more extensively use more references, and tend towards ``younger'' references, compared to authors who collaborate less. Here, we may hypothesise that more collaborative articles (i.e.~those not solely involving researchers at DEAL institutions) may receive more input into the writing and referencing process from researchers that are not subjected to the DEAL access restrictions, and thus any measurable effect on citing behaviour should be weakened in the more collaborative groups. Our analysis, where articles are divided with respect to the four previously defined collaboration classes (``Single author'', ``DEAL collaboration'', ``National collaboration'' and ``International collaboration'') is shown in Figure \ref{fig:references-publisher-year-collaboration}. Overall citation patterns over time are at least partially reflective of publishing patterns (Figure \ref{fig:items-publisher-year-collaboration}), e.g.~the lowest number of citations are found in single-author and national collaborations, where the number of publications is also lowest. However, we also observe some differences between groups that would appear to be decoupled from publication volume alone - for example, articles published by single authors have a lower proportion of articles that cite at least a single Elsevier article (38\%, aggregated over all years) compared to articles published from DEAL collaborations (56\%), national collaborations (65\%) or international collaborations (62\%). Interestingly, in terms of the total proportion of citations made to Elsevier articles, all four groups are relatively closer, with 22\% of citations from single authors, 25\% from DEAL collaborations, 25\% from national collaborations and 23\% from international collaborations being made to Elsevier articles (aggregated over all years).

In terms of variation over time, in a previous section we noted the strong decrease in Elsevier's market share of single-author articles from 2012 to 2020 (from 20.8\% to 14.2\%), compared to the more homogeneous publication behaviour of the three collaborative groups (Figure \ref{fig:items-publisher-year-collaboration}), which all displayed reduced proportions of articles published in Elsevier journals from 2016 to 2020. The evolution of citing behaviour, however, does not appear to follow the same patterns - in particular for single-author papers, we observe a long-term increase in the proportion of citations made to Elsevier articles (from 20.7\% in 2012 to 21.9\% in 2020), opposite to the trends found in publishing behaviour for the same group. Articles published by DEAL collaborations cited proportionally more Elsevier articles year-on-year between 2012 and 2018, and subsequently less in 2019 and 2020 (including a -1.1\% YOY market share change in 2020). The picture from the national and international collaboration groups is less consistent, e.g.~the proportion of citations to articles in Elsevier journals from national collaborations fluctuated in several years by \textgreater1\%, but without a clear trend over time, and variations in international collaborations tend to be less pronounced, with the exception of an increase of 1.3\% in 2018. changes in citing behaviour resulting from Elsevier access restrictions do not appear to reflect similar dynamics of publishing behaviour between different collaboration groups: of all groups, only articles published by DEAL collaborations show a decrease in the proportion of Elsevier citations concomitant with access restrictions beginning in 2018, whilst in other groups we do not find strong evidence for consistent changes in citing behaviour of Elsevier articles over time.

\hypertarget{how-did-citing-behaviour-vary-with-respect-to-oa-status}{%
\paragraph{How did citing behaviour vary with respect to OA status?}\label{how-did-citing-behaviour-vary-with-respect-to-oa-status}}

\begin{figure}
\centering
\includegraphics{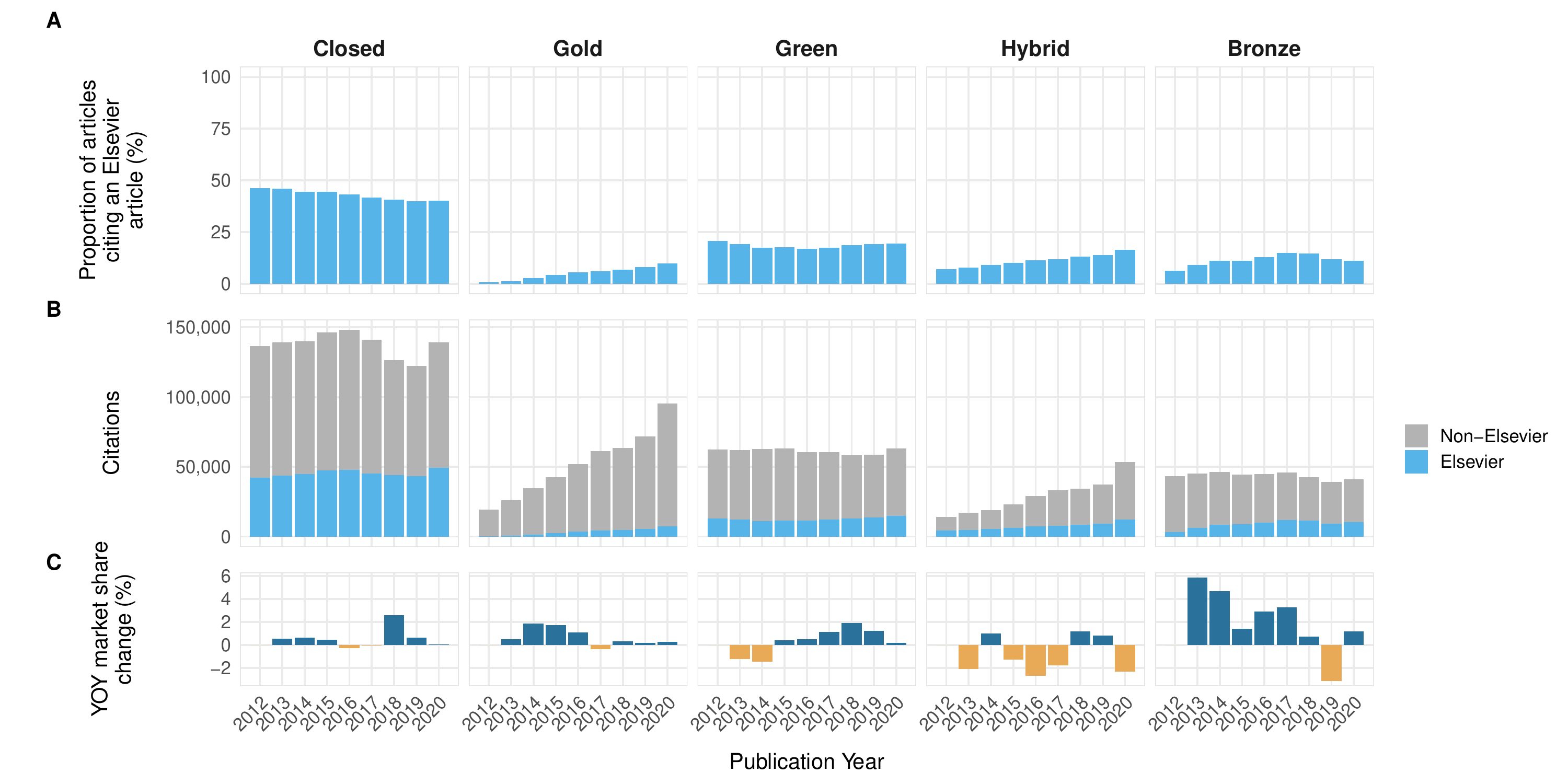}
\caption{\label{fig:references-publisher-year-oa}Citing behaviour of DEAL researchers, 2012-2020, dependent on OA status of the cited article. All data is based on citations from articles published by DEAL researchers to articles with a citation age of 2 years or less. (A) Proportion of articles published by DEAL researchers that cite at least one single Elsevier article. (B) Total number of citations from articles published DEAL researchers to articles published in Elsevier versus non-Elsevier journals. (C) Year-on-year (YOY) change in Elsevier's market share of citations from articles published by DEAL researchers.}
\end{figure}

In a previous section we analysed the publishing behaviour of researchers at DEAL institutions with respect to the OA status of the articles that they published. Here, we also consider the citing behaviour of researchers at DEAL institutions with respect to OA status, with the difference that OA status refers to that of the \emph{cited} article rather than the \emph{published} article; that is to say, we aim to determine whether researchers preferentially cite OA (or certain types of OA) articles over closed articles, and how these patterns have changed over time with respect to articles published in Elsevier journals. Elsevier access restrictions may have hindered the ability of researchers to access (and therefore cite) articles, however, OA articles would have remained accessible. In this case, a reasonable expectation would be that the proportion of citations to OA articles in Elsevier journals would have been greater after 2018 compared to those of closed articles. Main results are shown in Figure \ref{fig:references-publisher-year-oa}.

In Figure \ref{fig:references-publisher-year-oa}A, we show the proportion of articles that cite an Elsevier article, as a function of OA type of the cited article; as an example, we observe that 46.2\% of all articles published by DEAL researchers in 2012 also cited at least a single Closed article in an Elsevier journal. Over time, the proportion of articles that cited a Closed article in an Elsevier journal decreased, with a final proportion of 40.2\% in 2020. Comparatively, the proportion of articles that cited a Gold or Hybrid article in an Elsevier journal increased over the same time period (Gold from 0.8\% to 9.9\%, Hybrid from 7.2\% to 16.4\%). Figure \ref{fig:references-publisher-year-oa}B shows the number of references made to different types of OA between 2012 and 2020, and the proportion of which were made to articles in Elsevier journals. In general, the number of references made to Closed articles has remained relatively static over time, whilst the number of references made to Gold and Hybrid OA has rapidly increased.

Both of these results (i.e.~differences in the proportion of articles citing a single Elsevier articles, and changes in the total number of citations) need to be placed in the context of general background OA publishing trends over time. For example, whilst we observe growth in the proportion of articles citing a Gold or Hybrid OA article, the number of Gold and Hybrid OA articles has grown over the same time period (Figure \ref{fig:items-publisher-year-oa}B), and thus it is difficult to attribute such trends to true changes in the citing preference of DEAL researchers (i.e.~that researchers may consciously cite more OA articles than Closed articles). However, what is clear is restricted access to Elsevier journals in 2018 did not have a disproportionate effect on the ability of researchers to cite closed articles in Elsevier journals compared to previously; in fact, after 2018 we observe YOY increases in the proportion of all citations to closed articles, that were made to articles in Elsevier journals (Figure \ref{fig:references-publisher-year-oa}C), suggesting that there were no stronger barriers to citing Elsevier articles than at any other publisher.

\hypertarget{discussion-and-conclusions}{%
\subsection{Discussion and Conclusions}\label{discussion-and-conclusions}}

In this study, we have assessed changes in publishing and citing behaviour of researchers at DEAL institutions in Germany, with the aim to investigate the effect of access restrictions to Elsevier journals in 2018.

In terms of publishing behaviour, we found that Elsevier's market share of articles published by DEAL researchers has fallen by from a peak of 25.3\% in 2015 to 20.6\% in 2020 (corresponding to an absolute decrease of \textasciitilde1,000 articles per year published in Elsevier journals). Although the beginning of this period of market share losses does not directly correspond to the timing of access restrictions in 2018, we noted conspicuously large YOY market share losses (in excess of 1\% per year) in 2019 and 2020. We analysed these changes with respect to the timing of contract cancellations for individual DEAL institutions, research disciplines, collaboration patterns and article OA status. In broad terms, we find relatively weak evidence for any differences in publishing patterns depending on the timing of contract cancellations, or by collaboration status, although the proportion of single-author articles published in Elsevier journals has strongly decreased over the past 8 years (not limited to the period following Elsevier access restrictions). In terms of research disciplines, we observed an overall tendency towards a long-term decrease in Elsevier's market share for 18 of the 22 disciplines studied, but a high degree of variation exists over time and between individual disciplines. This may partly be the result of including a range of disciplines where publishing timelines strongly differ.

A particularly interesting feature of our results relates to that of OA publishing. Whilst OA has rapidly grown in Germany in recent years, OA growth at Elsevier has been comparatively slow - only 36.8\% of all articles published by DEAL researchers in Elsevier journals in 2020 were openly available (either at the journal page, or in a Green OA repository), compared to 68.8\% published in all journals. In the last 2-3 years in particular, Elsevier has lost considerable proportions of the OA market, likely driven by the successful conclusion of agreements between DEAL and other publishers (e.g.~Springer Nature and Wiley; \href{https://www.cesifo.org/en/publikationen/2021/working-paper/impact-german-deal-competition-academic-publishing-market}{Haucap et al., 2021}) and transfer of articles from DEAL researchers to those venues. Journals from the born-OA publishers Frontiers and MPDI have also gained popularity among authors affiliated with DEAL institutions in recent years (Figure \ref{fig:items-overview}A), which may have acted to redistribute articles from Elsevier journals to a wider range of competitors.

In the context of previous surveys conducted in response to Elsevier access restrictions our results show a surprisingly moderate response of the publishing behaviour of DEAL researchers. Of 384 researchers surveyed at the Faculty of Medicine of the University of Münster, 29\% reported that they would no longer write or review articles for Elsevier journals\footnote{Archived webpage: \url{https://web.archive.org/web/20210429122105/https://www.uni-muenster.de/ZBMed/aktuelles/27850}}, whilst 51\% of respondents to a survey by in the Bibsam consortium (in response to a similar situation in Sweden) reported that they were negatively affected in their desire to publish with Elsevier (\href{http://doi.org/10.1629/uksg.507}{Olson et al., 2020}). In contrast, our results show only a total decrease of Elsevier's market share of articles by DEAL researchers of \textasciitilde2.6\% in 2019 and 2020. Continued monitoring of the situation over the following years will provide more evidence as to the long-term effect on the motivation of DEAL researchers to publish in Elsevier journals in light of continued access restrictions (or alternatively, the removal of access restrictions if future negotiations result in a successful publishing agreement between DEAL and Elsevier) and the development of new publishing options at other publishers.

In terms of citing behaviour, we found that researchers have cited proportionally less Elsevier articles after 2018 than prior to 2018, but the effect is small in comparison to that observed for publishing behaviour: overall, Elsevier's share of references (limited to ``newer'' references with a maximum 2-year citation age) only fell from 24.9\% in 2018 to 23.9\% in 2020. Such effects do not seem to imply a markedly reduced ability of researchers to cite Elsevier articles after access restrictions came into place in 2018. As with publishing behaviour, we investigated these effects with respect to respect to the timing of contract cancellations for individual DEAL institutions, research disciplines, collaboration patterns and article OA status. However, whilst all of these units of analysis broadly agreed with the overall results, we found a lack of consistent patterns that can clearly by attributed to the implementation of Elsevier access restrictions in 2018. With respect to OA citing behaviour, where we previously noted some large-scale changes in publishing behaviour, we found that DEAL researchers are citing closed access articles in similar proportions to before the access restrictions were implemented, suggesting that no stronger barriers to citation exist for Elsevier journals compared to other publishers.

An obvious question that arises from these results, is that if researchers do not have access to Elsevier articles, why are they still able to cite them in such large volumes? A logical expectation would be that if authors are unable to access an article, they should not be able to read it and therefore not be able to cite it. This expectation is also rooted in the large number of studies over the past two decades showing a ``citation advantage'' of OA articles over non-OA articles (c.f. \href{https://doi.org/10.7717/peerj.4375}{Piwowar et al., 2018}), implying that accessibility and ability to cite an article are causally linked. This question may therefore be answered by considering two related mechanisms: firstly in how researchers gain access to articles, and secondly how they actually cite articles in practice.

With respect to the first mechanism, authors possess a number of strategies to access articles beyond institutional subscriptions. Sharing articles directly within informal networks of colleagues (e.g.~via email) remains a permitted practice (see, for example, Elsevier's guidance page on ``\href{https://www.elsevier.com/authors/submit-your-paper/sharing-and-promoting-your-article}{Sharing and promoting your article}''). Some researchers have taken this a step further by requesting articles directly from their networks of followers on various social media websites (e.g.~by posting an article request containing the \#icanhazpdf hashtag on Twitter; \href{https://doi.org/10.5596/c16-009}{Swab and Romme, 2016}). Interlibrary loans (ILL), whereby a library may borrow an article from the collection of another library, and Document Delivery Services (e.g.~\href{https://www.subito-doc.de/?lang=en}{Subito}) are another option through which researchers may request articles from their institutional libraries. However, a previous analysis on the effect of ``Big Deal'' cancellations has found that changes in the volume of ILL requests following such cancellations were usually small (\href{https://arxiv.org/abs/2009.04287}{Simard et al., 2020}). Interestingly, in the previously-discussed survey of the Bibsam consortium in Sweden, 23\% of respondents answered that they received access to articles through their library when denied access, yet data on article delivery services reported no increases in ILL requests for nine months following their contract cancellation (\href{http://doi.org/10.1629/uksg.507}{Olson et al., 2020}).

\begin{figure}

{\centering \includegraphics[width=0.75\linewidth]{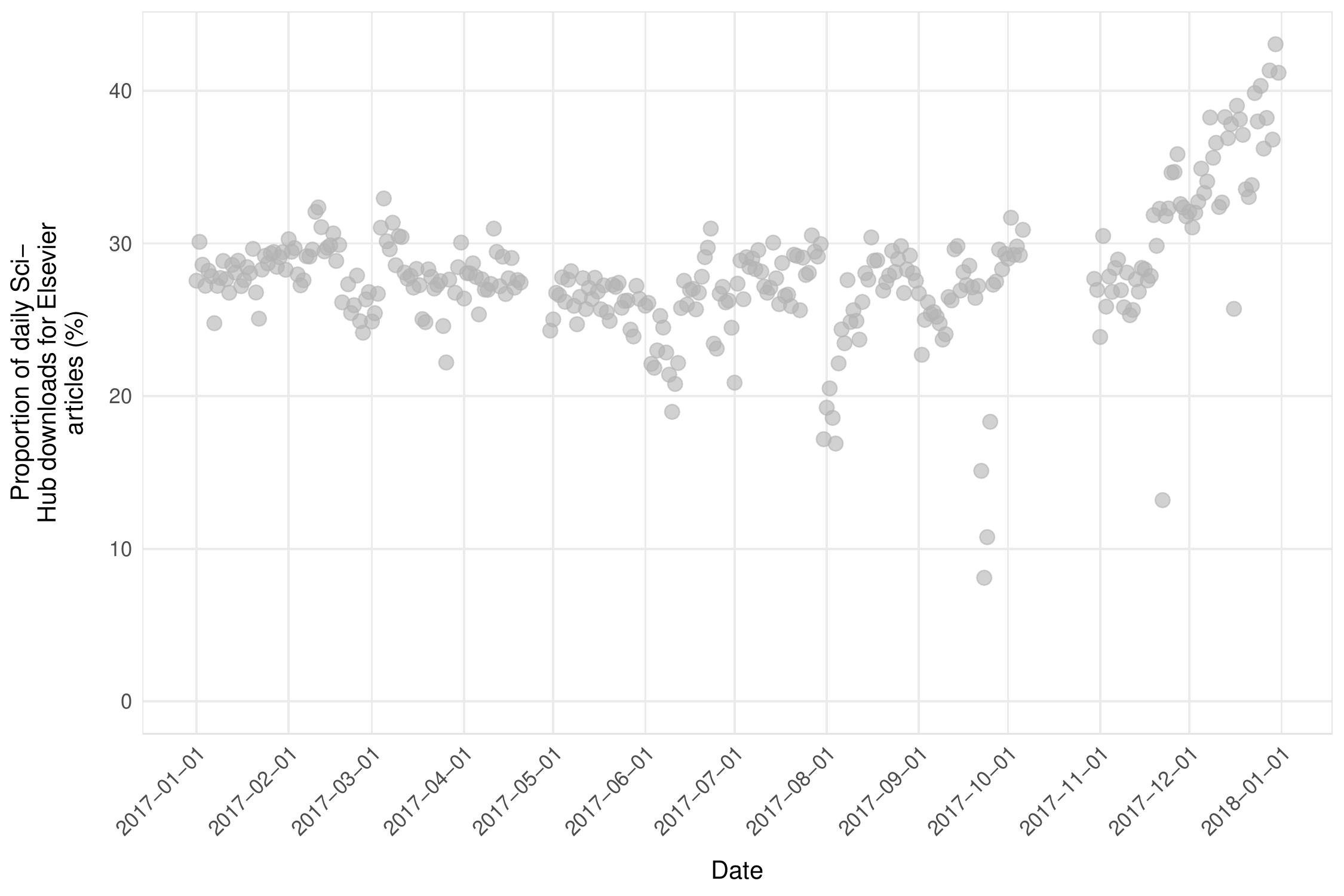} 

}

\caption{Proportion of daily Sci-Hub downloads from Germany in 2017 for Elsevier articles. Data from \href{http://doi.org/10.5281/zenodo.1286284}{Strecker, (2018)}.}\label{fig:scihub-germany}
\end{figure}

Other methods that researchers may use to access articles involve so-called ``shadow libraries'' or ``pirate OA'', where articles are made available on the web with disregard to any existing copyright. Two such venues have gained wide-scale popularity in recent years: \href{https://sci-hub.se/}{Sci-Hub} and \href{https://www.researchgate.net/}{ResearchGate}. Of the respondents to the Bibsam survey, 26\% answered that when they cannot access an article they instead sought access on ResearchGate, and 14\% on Sci-Hub (\href{http://doi.org/10.1629/uksg.507}{Olson et al., 2020}). A similar survey of researchers at the Faculty of Medicine of the University of Münster found that 46\% of researchers sought articles on ResearchGate when access was unavailable at their university\footnote{Archived webpage: \url{https://web.archive.org/web/20210429130228/https://www.uni-muenster.de/ZBMed/aktuelles/27987}}. A previous study has found that more than 50\% of articles on ResearchGate are in infringement of copyright and publishers' policies (\href{https://doi.org/10.1007/s11192-017-2291-4}{Jamali, 2017}), which led to major publishers including Elsevier to take legal action against ResearchGate in 2018 (\href{https://doi.org/10.1038/d41586-018-06945-6}{Else, 2018}). Sci-Hub, founded by Alexandra Elbakyan in 2011, is another shadow library with large-scale coverage: an analysis in 2018 found that it provides access to nearly all available scholarly literature (\href{https://doi.org/10.7554/eLife.32822}{Himmelstein et al, 2018}). In Figure \ref{fig:scihub-germany} we analysed the rates of daily Sci-Hub downloads from Germany in 2017, using freely available and geo-coded access logs that were released by Sci-Hub (data for Germany was previously aggregated by \href{http://doi.org/10.5281/zenodo.1286284}{Strecker (2018)}). The figure shows the proportion of all downloads that were made for Elsevier articles each day. To our knowledge, no more recent data is available beyond the end of 2017, thus we cannot measure changes in download rates following Elsevier access restrictions in 2018; however, our data do cover the brief 6-week period that Elsevier restricted access at the beginning of 2017, during which we observe no increase in the proportion of downloads made to Elsevier articles compared to articles from other venues. Interestingly, the proportion of downloads for Elsevier articles increased dramatically by \textasciitilde10\% in December 2017 compared to previous months, though it is not clear what drove these increased download rates.

With respect to researchers citation practices, and how these may influence the continually high citation rates of ``inaccessible'' Elsevier articles, a more contentious hypothesis is that scientists do not necessarily read the full-text articles before citing them (i.e.~they read only the abstracts, or do not read them at all), making the issue of access obsolete. Previous studies on the frequency and patterns of misprints in reference lists have concluded that 70-90\% of references are simply copied from other articles' reference lists (\href{https://doi.org/10.1007/s11192-005-0028-2}{Simkin and Roychowdhury, 2005}), with estimates that only \textasciitilde20\% of all authors that cite a paper have actually read it (\href{https://doi.org/10.1002/asi.20653}{Simkin and Roychowdhury, 2007}). We cannot determine whether this is the case for researchers in Germany, but this may present an interesting avenue for future studies that also look at differences in citation rates between OA and non-OA articles.

In summary, our aim for this study was to provide first insights into the direct effects of restricted access to a set of journals on researchers' publishing and citing behaviours, from a bibliometric perspective. The results may be important for a range of stakeholders, both within and outside of Germany, involved in the negotiations of publishing agreements with commercial publishers. We hope that the methodology presented can be used as a blueprint for follow-up studies and/or long-term monitoring of these effects over a longer period of time since we assume that only recurring studies will be able to reveal the true impact of the DEAL restrictions. Our study has also shown, that - in addition to surveys that capture the perception of researchers and that rely on self-reported data - there is a strong need for the analysis of behavioural data to gain a holistic view on researchers' publication and citation processes.

\hypertarget{limitations-and-future-directions}{%
\subsection{Limitations and Future Directions}\label{limitations-and-future-directions}}

Our study has several limitations, which may be discussed and improved upon in future studies.

Firstly, we rely heavily on article, author and institutional data from a single bibliometric database, Dimensions. Future studies may test the robustness of our results by comparing them against results obtained through other bibliometric data sources, for example through \href{https://www.webofknowledge.com/}{Web of Science} or \href{https://www.scopus.com}{Scopus}. These databases may have additional advantages in the quality and coverage of author role information (e.g.~more complete corresponding author information) and article types.

A second important limitation of this study is that we consider changes in publishing and citing behaviour over time with respect to publishing timelines, rather than submission timelines. Articles spend a significant proportion of time following submission in peer-review and publication cycles, and these timelines vary strongly by discipline; disciplines such as Business or Economics take 18 months on average between submission and publication (\href{https://doi.org/10.1016/j.joi.2013.09.001}{Björk and Solomon, 2013}). Given that our analysis period covers at maximum a period of 30-months (i.e.~to the end of December 2020) following the introduction of Elsevier access restrictions in July 2018, we are likely only capturing the early effects on researchers' behaviour. Future studies should therefore monitor these effects over longer periods; future negotiations and any potential publishing agreements made with Elsevier may also complicate such analysis further.

This study used bibliometric data to investigate potential effects of journal cancellations. However, information about the OA business model and revenue streams become more and more important in order to assess national licensing activities of libraries aiming for OA. In the case of Elsevier, it is particularly interesting to know whether authors can make use of grants to pay for publication fees. Another question is whether the open access publication would be considered as compliant by the DEAL consortium, because Elsevier publishes OA journals, which publish proceedings and are not listed in the Directory of Open Access Journals (DOAJ). Likewise, a considerable proportion of articles are made available through Elsevier's Open Archive Program after an embargo period. Future studies could make use of journal lists from related agreements between national consortia and Elsevier, as well as publisher-provided invoicing data (\href{http://arxiv.org/abs/2102.04789\%7D}{Jahn et al., 2021}), for an in-depth analysis of OA publication activities and potential funding sources.

Finally, in this study we have taken a descriptive, quantitative approach to understanding changes in researchers publishing and citing behaviour over time. We rely solely on large-scale bibliometric data, and have not conducted any statistical modelling (e.g.~to test for the interrelation between various factors that may influence publishing or citing rates) nor attempted to qualify these findings through other more qualitative methods (e.g.~surveys, interviews with researchers, librarians and other stakeholders) that would be needed to understand the exact underlying mechanisms driving these changes. It is particularly interesting, that we were able to detect a strong contradiction between self-reported data and bibliometric data on the effect of DEAL on publishing and citing. So, we encourage such follow-up studies, particularly those that use a multi-method approach to explore researchers' knowledge of Elsevier access restrictions, its effect on their day-to-day research activities, and their associated motivations for publishing (or not publishing) in Elsevier journals in future.

\hypertarget{references}{%
\subsection{References}\label{references}}

\hypertarget{refs}{}
\begin{CSLReferences}{1}{0}
\leavevmode\hypertarget{ref-bjork_publishing_2013}{}%
Björk, B.-C., \& Solomon, D. (2013). The publishing delay in scholarly peer-reviewed journals. \emph{Journal of Informetrics}, \emph{7}(4), 914--923. \url{https://doi.org/10.1016/j.joi.2013.09.001}

\leavevmode\hypertarget{ref-bornmann_field_2018}{}%
Bornmann, L. (2018). Field classification of publications in {Dimensions}: A first case study testing its reliability and validity. \emph{Scientometrics}, \emph{117}(1), 637--640. \url{https://doi.org/10.1007/s11192-018-2855-y}

\leavevmode\hypertarget{ref-brand_beyond_2015}{}%
Brand, A., Allen, L., Altman, M., Hlava, M., \& Scott, J. (2015). Beyond authorship: Attribution, contribution, collaboration, and credit. \emph{Learned Publishing}, \emph{28}(2), 151--155. \url{https://doi.org/10.1087/20150211}

\leavevmode\hypertarget{ref-chamberlain_rcrossref_2020}{}%
Chamberlain, S., Zhu, H., Jahn, N., Boettiger, C., \& Ram, K. (2020). \emph{Rcrossref: {Client} for {Various} '{CrossRef}' '{APIs}'}. \url{https://CRAN.R-project.org/package=rcrossref}

\leavevmode\hypertarget{ref-crossref_january_2021}{}%
Crossref. (2021). \emph{January 2021 {Public} {Data} {File} from {Crossref}}. \url{https://doi.org/10.13003/GU3DQMJVG4}

\leavevmode\hypertarget{ref-else_dutch_2018}{}%
Else, H. (2018a). Dutch publishing giant cuts off researchers in {Germany} and {Sweden}. \emph{Nature}. \url{https://doi.org/10.1038/d41586-018-05754-1}

\leavevmode\hypertarget{ref-else_major_2018}{}%
Else, H. (2018b). Major publishers sue {ResearchGate} over copyright infringement. \emph{Nature}. \url{https://doi.org/10.1038/d41586-018-06945-6}

\leavevmode\hypertarget{ref-haucap_impact_2021}{}%
Haucap, J., Moshgbar, N., \& Wolfgang Benedikt, S. (2021). The {Impact} of the {German} '{DEAL}' on {Competition} in the {Academic} {Publishing} {Market}. \emph{CESifo}, \emph{Working Paper No. 8963}. \url{https://www.cesifo.org/en/publikationen/2021/working-paper/impact-german-deal-competition-academic-publishing-market}

\leavevmode\hypertarget{ref-herzog_response_2018}{}%
Herzog, C., \& Lunn, B. K. (2018). Response to the letter {``{Field} classification of publications in {Dimensions}: A first case study testing its reliability and validity.''} \emph{Scientometrics}, \emph{117}(1), 641--645. \url{https://doi.org/10.1007/s11192-018-2854-z}

\leavevmode\hypertarget{ref-heyman_cost_2016}{}%
Heyman, T., Moors, P., \& Storms, G. (2016). On the {Cost} of {Knowledge}: {Evaluating} the {Boycott} against {Elsevier}. \emph{Frontiers in Research Metrics and Analytics}, \emph{1}. \url{https://doi.org/10.3389/frma.2016.00007}

\leavevmode\hypertarget{ref-himmelstein_sci-hub_2018}{}%
Himmelstein, D. S., Romero, A. R., Levernier, J. G., Munro, T. A., McLaughlin, S. R., Greshake Tzovaras, B., \& Greene, C. S. (2018). Sci-{Hub} provides access to nearly all scholarly literature. \emph{eLife}, \emph{7}, e32822. \url{https://doi.org/10.7554/eLife.32822}

\leavevmode\hypertarget{ref-hobert_entwicklung_2021}{}%
Hobert, A., Haupka, N., \& Najko, N. (2021). Entwicklung und {Typologie} des {Datendiensts} {Unpaywall}. \emph{BIBLIOTHEK -- Forschung Und Praxis}. \url{https://doi.org/10.18452/22728}

\leavevmode\hypertarget{ref-hobert_open_2021}{}%
Hobert, A., Jahn, N., Mayr, P., Schmidt, B., \& Taubert, N. (2021). Open access uptake in {Germany} 2010--2018: Adoption in a diverse research landscape. \emph{Scientometrics}. \url{https://doi.org/10.1007/s11192-021-04002-0}

\leavevmode\hypertarget{ref-jahn_transparency_2021}{}%
Jahn, N., Matthias, L., \& Laakso, M. (2021). Transparency to hybrid open access through publisher-provided metadata: {An} article-level study of {Elsevier}. \emph{arXiv:2102.04789 {[}cs{]}}. \url{http://arxiv.org/abs/2102.04789}

\leavevmode\hypertarget{ref-jamali_copyright_2017}{}%
Jamali, H. R. (2017). Copyright compliance and infringement in {ResearchGate} full-text journal articles. \emph{Scientometrics}, \emph{112}(1), 241--254. \url{https://doi.org/10.1007/s11192-017-2291-4}

\leavevmode\hypertarget{ref-kieselbach_projekt_2020}{}%
Kieselbach, S. (2020). \emph{Projekt {DEAL} -- {Springer} {Nature} {Publish} and {Read} {Agreement}}. \url{https://doi.org/10.17617/2.3174351}

\leavevmode\hypertarget{ref-kwiek_what_2020}{}%
Kwiek, M. (2020). What large-scale publication and citation data tell us about international research collaboration in {Europe}: Changing national patterns in global contexts. \emph{Studies in Higher Education}, 1--21. \url{https://doi.org/10.1080/03075079.2020.1749254}

\leavevmode\hypertarget{ref-milojevic_how_2012}{}%
Milojević, S. (2012). How {Are} {Academic} {Age}, {Productivity} and {Collaboration} {Related} to {Citing} {Behavior} of {Researchers}? \emph{PLoS ONE}, \emph{7}(11), e49176. \url{https://doi.org/10.1371/journal.pone.0049176}

\leavevmode\hypertarget{ref-olsson_cancelling_2020}{}%
Olsson, L., Lindelöw, C. H., Österlund, L., \& Jakobsson, F. (2020). Cancelling with the world's largest scholarly publisher: Lessons from the {Swedish} experience of having no access to {Elsevier}. \emph{Insights the UKSG Journal}, \emph{33}, 13. \url{https://doi.org/10.1629/uksg.507}

\leavevmode\hypertarget{ref-parolo_attention_2015}{}%
Parolo, P. D. B., Pan, R. K., Ghosh, R., Huberman, B. A., Kaski, K., \& Fortunato, S. (2015). Attention decay in science. \emph{Journal of Informetrics}, \emph{9}(4), 734--745. \url{https://doi.org/10.1016/j.joi.2015.07.006}

\leavevmode\hypertarget{ref-r_core_team_r_2020}{}%
R Core Team. (2020). \emph{R: {A} {Language} and {Environment} for {Statistical} {Computing}}. R Foundation for Statistical Computing. \url{https://www.R-project.org/}

\leavevmode\hypertarget{ref-r_special_interest_group_on_databases_r-sig-db_dbi_2021}{}%
R Special Interest Group on Databases (R-SIG-DB), Wickham, H., \& Müller, K. (2021). \emph{{DBI}: {R} {Database} {Interface}}. \url{https://CRAN.R-project.org/package=DBI}

\leavevmode\hypertarget{ref-sander_projekt_2019}{}%
Sander, F., Hermann, G., Hippler, H., Meijer, G., \& Schimmer, R. (2019). \emph{Projekt {DEAL} -- {John} {Wiley} \&amp; {Son} {Publish} and {Read} {Agreement}}. \url{https://doi.org/10.17617/2.3027595}

\leavevmode\hypertarget{ref-sanchez-gil_reference_2018}{}%
Sánchez-Gil, S., Gorraiz, J., \& Melero-Fuentes, D. (2018). Reference density trends in the major disciplines. \emph{Journal of Informetrics}, \emph{12}(1), 42--58. \url{https://doi.org/10.1016/j.joi.2017.11.003}

\leavevmode\hypertarget{ref-simard_aftermath_2020}{}%
Simard, M.-A., Priem, J., \& Piwowar, H. (2020). The aftermath of {Big} {Deal} cancellations and their impact on interlibrary loans. \emph{arXiv:2009.04287 {[}cs{]}}. \url{http://arxiv.org/abs/2009.04287}

\leavevmode\hypertarget{ref-simkin_stochastic_2005}{}%
Simkin, M. V., \& Roychowdhury, V. P. (2005). Stochastic modeling of citation slips. \emph{Scientometrics}, \emph{62}(3), 367--384. \url{https://doi.org/10.1007/s11192-005-0028-2}

\leavevmode\hypertarget{ref-simkin_mathematical_2007}{}%
Simkin, Mikhail V., \& Roychowdhury, V. P. (2007). A mathematical theory of citing. \emph{Journal of the American Society for Information Science and Technology}, \emph{58}(11), 1661--1673. \url{https://doi.org/10.1002/asi.20653}

\leavevmode\hypertarget{ref-strecker_sci-hub_2018}{}%
Strecker, D. (2018). \emph{Sci-{Hub} {Downloads} {From} {Germany}}. Zenodo. \url{https://doi.org/10.5281/ZENODO.1286284}

\leavevmode\hypertarget{ref-swab_scholarly_2016}{}%
Swab, M., \& Romme, K. (2016). Scholarly {Sharing} via {Twitter}: \#icanhazpdf {Requests} for {Health} {Sciences} {Literature}. \emph{Journal of the Canadian Health Libraries Association}, \emph{37}(1). \url{https://doi.org/10.5596/c16-009}

\leavevmode\hypertarget{ref-vogel_german_2017}{}%
Vogel, G. (2017a). German researchers start 2017 without {Elsevier} journals. \emph{Science}, \emph{355}(6320), 17--17. \url{https://doi.org/10.1126/science.355.6320.17}

\leavevmode\hypertarget{ref-vogel_elsevier_2017}{}%
Vogel, G. (2017b). Elsevier journals are back online at 60 {German} institutions that had lost access. \emph{Science}. \url{https://doi.org/10.1126/science.aal0753}

\leavevmode\hypertarget{ref-vogel_german_2017-1}{}%
Vogel, G. (2017c). German researchers resign from {Elsevier} journals in push for nationwide open access. \emph{Science}. \url{https://doi.org/10.1126/science.aar2142}

\leavevmode\hypertarget{ref-wickham_welcome_2019}{}%
Wickham, H., Averick, M., Bryan, J., Chang, W., McGowan, L. D., François, R., Grolemund, G., Hayes, A., Henry, L., Hester, J., Kuhn, M., Pedersen, T. L., Miller, E., Bache, S. M., Müller, K., Ooms, J., Robinson, D., Seidel, D. P., Spinu, V., \ldots{} Yutani, H. (2019). Welcome to the tidyverse. \emph{Journal of Open Source Software}, \emph{4}(43), 1686. \url{https://doi.org/10.21105/joss.01686}

\leavevmode\hypertarget{ref-wickham_bigrquery_2020}{}%
Wickham, H., \& Bryan, J. (2020). \emph{Bigrquery: {An} {Interface} to {Google}'s '{BigQuery}' '{API}'}. \url{https://CRAN.R-project.org/package=bigrquery}

\end{CSLReferences}

\end{document}